\documentclass[%
 reprint,
superscriptaddress,
 amsmath,amssymb,
 aps,
pra,
]{revtex4-2}

\usepackage{amsmath}
\usepackage{amssymb}
\usepackage{amsthm}
\usepackage{physics}
\usepackage{graphicx}% Include figure files
\usepackage{dcolumn}% Align table columns on decimal point
\usepackage{bm}% bold math
\usepackage{hyperref}% add hypertext capabilities
\theoremstyle{definition}

\usepackage{xcolor}

\begin{document}

\preprint{APS/123-QED}

\title{Multipartite device-independent quantum key distribution using \textit{W} states}
% \title{Long-Distance Continuous-Variable Quantum Key Distribution\\ with Highly Non-Gaussian State}% Force line breaks with \\

\author{Makoto Ishihara}
 \email{llmakomako.arg1076@keio.jp}
 \affiliation{%
 Department of Electronics and Electrical Engineering, Keio University, 3-14-1 Hiyoshi, Kohoku-ku, Yokohama 223-8522, Japan
}%
\author{Wojciech Roga}%
 \email{wojciech.roga@keio.jp}
 \affiliation{%
 Department of Electronics and Electrical Engineering, Keio University, 3-14-1 Hiyoshi, Kohoku-ku, Yokohama 223-8522, Japan
}%
\author{Masahiro Takeoka}%
 \email{takeoka@elec.keio.ac.jp}
\affiliation{%
 Department of Electronics and Electrical Engineering, Keio University, 3-14-1 Hiyoshi, Kohoku-ku, Yokohama 223-8522, Japan
}%

\date{\today}% It is always \today, today,

\begin{abstract}

Multipartite device-independent quantum key distribution (DI-QKD), also known as device-independent conference key agreement, enables more than two remote parties to share a common key with information-theoretic security even without trusting the devices. So far, several multipartite DI-QKD protocols have been proposed where Greenberger-Horne-Zeilinger (GHZ) states are used as multipartite entanglement. A natural question is then whether one can construct multipartite DI-QKD with the other type of multipartite entanglement. \textit{W} state is of particular interest since it is intrinsically different from GHZ state and in some cases, easier to optically implement. In this paper, we show that multipartite DI-QKD is possible with \textit{W} states. To this end, we construct Bell inequalities largely violated by \textit{W} states, which can be used for the multipartite DI-QKD. Furthermore, we consider several different implementation scenarios. First, we analyze the minimum required detection efficiencies to extract finite amount of keys. Then we propose a long-distance multipartite DI-QKD protocol with single-photon interference and make detailed analyses with several physical implementation scenarios. We show that the protocol enables secret key distribution over longer distances than the existing multipartite DI-QKD protocols based on GHZ states. 
This study provides new insight about the relationship between multipartite entanglement and device-independent quantum information processing as well as opens an alternative path toward long-distance multipartite DI-QKD.

\end{abstract}

\maketitle

%\tableofcontents

\section{\label{sec:introduction}INTRODUCTION}
Quantum key distribution (QKD) is one of the most prominent quantum information technologies that enables two remote parties to share an information-theoretically secure secret key~\cite{Bennett2014, Ekert1991, BBM1992, Bennett1992}. After the seminal paper by Bennett and Brassard~\cite{Bennett2014}, many QKD protocols have been proposed and demonstrated experimentally~\cite{Xu2020, Pirandola2020}, e.g., testbeds for QKD implementation have been constructed around the world~\cite{Elliott2002, Elliott2005, Peev2009, Sasaki2011, Dynes2019, YAChen2021, TYChen2021, Martin2024}. While two remote parties share a secret key in QKD protocols, more than two parties share a common secret key in multipartite QKD protocols (sometimes referred as conference key agreement). As in QKD, multipartite QKD has been widely investigated both theoretically~\cite{Wu2016, Epping2017, Zhang2018, Grasselli2018, Ottaviani2019, Grasselli2019, Zhao2020, Carrara2023} and experimentally~\cite{Proietti2021, Pickston2023, Yang2024} (see also Ref.~\cite{Murta2020}).

The security of standard QKD or multipartite QKD protocols relies on the assumption that the devices that are used are completely characterized during implementation. However, it is challenging to satisfy this assumption in a practical scenario. To relax this requirement, measurement-device-independent protocols which do not require complete characterization of detectors have been proposed~\cite{Lo2012, Pirandola2015, Lucamarini2018}. Nevertheless, these protocols still require the assumption that devices for preparing quantum states must be perfectly characterized.

Device-independent QKD (DI-QKD) guarantees its security without any assumptions on the inner workings of the devices~\cite{Zapatero2023, Primaatmaja2023}. The security of DI-QKD protocols is certified by using the loophole-free violation of Bell inequalities~\cite{Acin2007, Pironio2009}. The violation of Bell inequalities forbids an eavesdropper, Eve, from performing eavesdropping strategies which do not admit the violation and guarantees the security of a shared key. Same as standard QKD protocols, several multipartite DI-QKD protocols have been proposed~\cite{Ribeiro2018, Holz2019, Ribeiro2019, Holz2020, Grasselli2023, Wooltorton2025, Ishihara2025}, where the Greenberger-Horne-Zeilinger (GHZ) states~\cite{Greenberger1990} are used as multipartite entanglement. 

%However, almost all multipartite DI-QKD protocols which have been proposed so far are based on the Greenberger-Horne-Zeilinger (GHZ) states~\cite{Greenberger1990}.

The \textit{W} state~\cite{Dur2000} is another kind of multipartite entangled state which is intrinsically different from the GHZ state. It is known that \textit{W} states have advantages in some quantum information technologies~\cite{MiguelRamiro2023}, e.g., \textit{W} states are more robust against losses compared to GHZ states and we can distribute \textit{W} states over long distances with high rates~\cite{Roga2023}. While research on Bell nonlocality using \textit{W} states has been widely conducted~\cite{Cabello2002, Sen2003, Heaney2011,  Brunner2012, Brask2012, Wang2013, Brask2013, Bancal2013, Sohbi2015, Barnea2015, Pal2015, Divianszky2016, Bjerrum2023}, it remains an open question whether \textit{W} states are useful for multipartite DI-QKD protocols. Realizing multipartite DI-QKD protocols based on \textit{W} states is not straightforward since, unlike with GHZ states, it is not possible to extract one bit of perfectly correlated key from one \textit{W} state. In other words, even without any experimental imperfections, a certain amount of key rate reduction at key distillation is inevitable for multipartite DI-QKD protocols with \textit{W} states. This makes the realization of \textit{W}-based DI-QKD challenging.

In this paper, we answer to the above open question affirmatively, i.e. we show that it is possible to construct multipartite DI-QKD with \textit{W} states. To this end, by using numerical optimization, we construct Bell inequalities whose violations with \textit{W} states provide large Eve's uncertainty on secret keys. For these Bell inequalities, we calculate multipartite DI-QKD key rates and show that multipartite DI-QKD protocols using \textit{W} states are possible.
Then, we consider the realistic conditions of its implementation. 
First, we analyze required detection efficiency to extract finite amount of keys when ideal {\it W} states are distributed among remote parties. 
Two different measurement modelings are examined: arbitrary Pauli measurement and measurement consisting of optical displacement operation followed by photon detection, which fits quantum memory-based detection and fully optical detection, respectively. 
Furthermore, we consider the scenario taking into account the channel losses to deliver {\it W} states to the remote parties. 
We propose the long-distance multipartite DI-QKD protocols with the efficient {\it W}-state distribution protocol developed in Ref.~\cite{Roga2023}. We show that multipartite DI-QKD with this protocol can distribute a secret key over longer distances compared to a multipartite DI-QKD protocol based on locally generated GHZ states~\cite{Ribeiro2018, Holz2019, Ribeiro2019}. We also consider implementation with Gaussian states. The optical entanglement generated by spontaneous parametric down conversion (SPDC) is known to be a Gaussian state and thus we examine the feasibility of a fully optical setup of our protocol.

The paper is structured as follows. In Sec. \ref{sec:protocol}, we describe our multipartite DI-QKD protocol using \textit{W} states. In Sec.~\ref{sec:keyrate}, we explain multipartite DI-QKD key rates and the difficulty of realizing multipartite DI-QKD protocols using \textit{W} states. In Sec.~\ref{sec:Bellineq}, we construct Bell inequalities which enable multipartite DI-QKD based on \textit{W} states. We also calculate classical and quantum limits of the Bell inequalities and show that \textit{W} states can largely violate those Bell inequalities. We analyze required detection efficiency to get positive multipartite DI-QKD key rates in Sec.~\ref{sec:required}. We apply the efficient \textit{W}-state distribution protocol to multipartite DI-QKD in Sec.~\ref{sec:roga}, and we analyze its experimental feasibility by considering Gaussian states and measurements in Sec.~\ref{sec:gaussian}. Finally, we conclude in Sec.~\ref{sec:conclusion}.

\section{Protocol}\label{sec:protocol}

\begin{figure*}[htbp]
 \centering
 \includegraphics[keepaspectratio, scale=0.4]{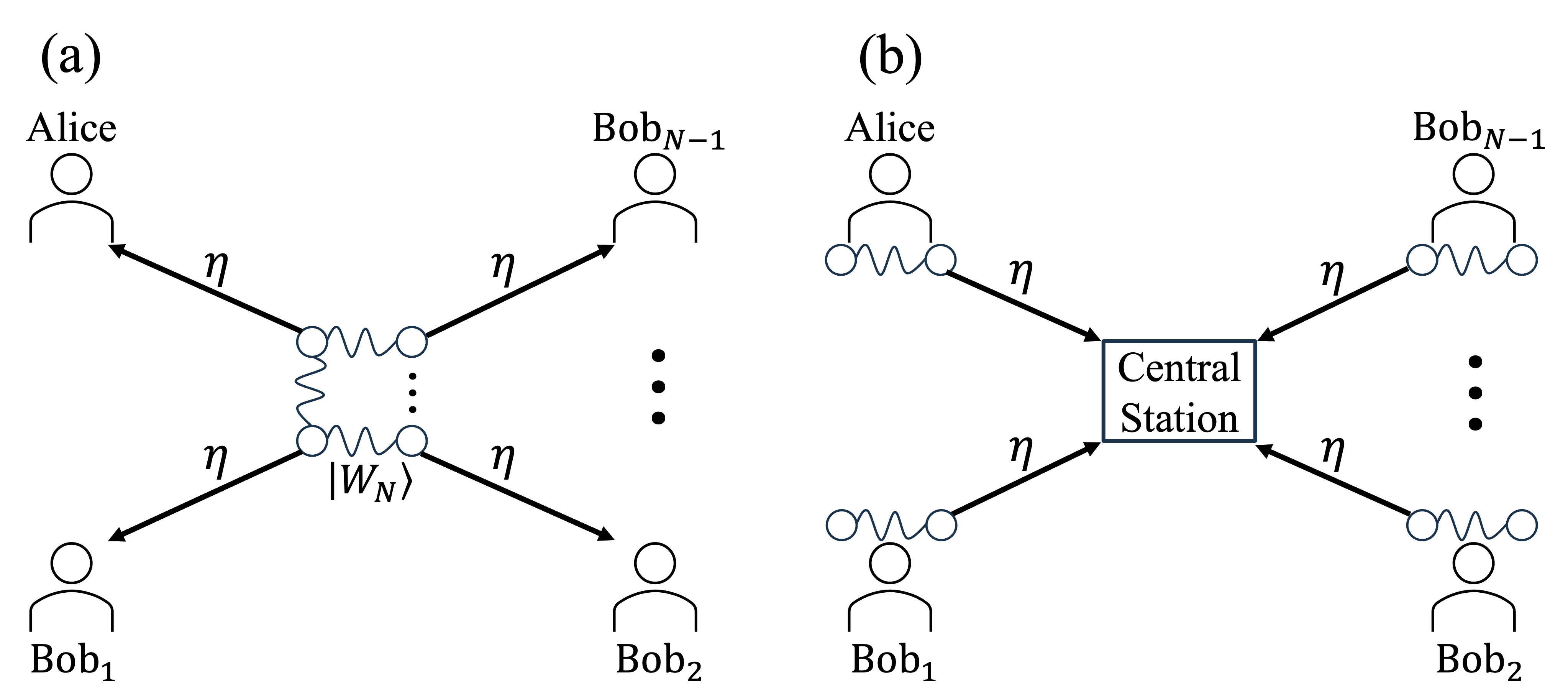}
 \caption{Schematics of multipartite DI-QKD using \textit{W} states. $N$ parties, Alice and $\text{Bob}_i$ ($i \in \{ 1, \ldots, N-1\}$) share a common secret key. A \textit{W} state is first distributed among the parties and the parties perform some measurements on the distributed \textit{W} state. The parties share a common secret key based on their measurement results. (a) Direct transmission. a \textit{W} state is generated locally and distributed among the parties through pure-loss channels with transmissivity $\eta$. (b) RIHT protocol. Each party locally prepares an entangled state and sends one part to a central station through a pure-loss channel with transmissivity $\eta$. A \textit{W} state is distributed among the parties when single-photon interference succeeds at the central station.}
 \label{fig:schematic}
\end{figure*}

We describe our multipartite DI-QKD protocol which is schematically shown in Fig.~\ref{fig:schematic} using \textit{W} states. We consider a $(N, m, d)$ scenario where $N$ parties, Alice and $\text{Bob}_i$ ($i \in \{ 1, \ldots, N-1 \}$) share a common secret key and each party has $m$ different measurement settings with $d$ outcomes. 

First, $N$ legitimate parties share an $N$-partite \textit{W} state $\ket{W_N}$
\begin{equation}
    \ket{W_N} = \frac{1}{\sqrt{N}} (\ket{00\cdots 01} + \ket{00\cdots 10} + \cdots + \ket{10 \cdots 00}),
\end{equation}
where $\ket{0}$ and $\ket{1}$ can be physically prepared with, e.g., horizontal and vertical polarization states or a vacuum and a single photon, respectively. We consider two different ways of distributing the \textit{W} state, that is, direct transmission and Roga-Ikuta-Horikiri-Takeoka (RIHT) protocol~\cite{Roga2023}. In the direct transmission, shown in Fig.~\ref{fig:schematic} (a), the \textit{W} state is locally generated and distributed among the parties. On the other hand, in the RIHT protocol, shown in Fig.~\ref{fig:schematic} (b), every party locally prepares a two-mode entangled state and sends one part to a central station. At the central station, single-photon interference is performed on these modes and the \textit{W} state is distributed among the parties if the interference measurement succeeds. We provide details on the RIHT protocol later, see Sec.~\ref{sec:roga}.

After the distribution of the \textit{W} state, the parties perform some measurements on the \textit{W} state. Let $M_{a|x}$ denote positive operator-valued measure (POVM) of Alice's measurement where $x \in \{0, \ldots, m-1 \}$ is measurement input and $a \in \{ 0, \ldots, d-1\}$ is measurement output. Also, for $i \in \{1, \ldots, N-1\}$, let $N_{b_i|y_i}$ denote POVM of $\text{Bob}_i$'s measurement where $y_i \in \{0, \ldots, m \}$ is measurement input and $b_i \in \{ 0, \ldots, d-1\}$ is measurement output. A part of events where Alice chooses $x = 0$ and $\text{Bob}_i \, (i \in \{1, \ldots, N-1 \})$ chooses $y_i=0$ is classified into key generation rounds to generate a common secret key. The rest of events are classified into test rounds for security testing. Let $p(a,b_1, \ldots, b_{N-1}|x, y_1, \ldots, y_{N-1})$ denote joint probability distribution of the legitimate parties for the test rounds. Furthermore, the parties perform the noisy preprocessing~\cite{Ho2020}. For the noisy preprocessing, Alice flips her key bits with probability $p_n$. This procedure decreases the correlation between Alice and the other parties, but it also decreases the correlation between Alice and Eve. Then, the net effect might be positive. Finally, the legitimate parties perform error correction and privacy amplification, sharing a secret key.

\section{DI-QKD key rate}\label{sec:keyrate}
Here, we describe key rates of our protocol. We focus on asymptotic key rates and we assume that the devices behave independently and identically during implementations. A multipartite DI-QKD key rate of our protocol is expressed as follows~\cite{Devetak2005}
\begin{equation}\label{eq:keyrate}
    r = P_\text{succ} (H(A|x^*, E) - \max_i H(A|B_i, x^*, y^*_i)),
\end{equation}
where $A$, $B_i$ and $E$ denote Alice's, $\text{Bob}_i$'s and Eve's system, respectively, $x^*$ and $y^*_i$ denote Alice's and $\text{Bob}_i$'s measurement inputs for the key generation rounds, and $P_\text{succ}$ is a probability that a \textit{W} state is successfully distributed among the parties. The first term is the conditional von Neumann entropy between Alice and Eve, corresponding to Eve's uncertainty about Alice's measurement outcomes. The second term shows the error correction cost between Alice and Bobs.

What makes the implementation of multipartite DI-QKD protocols using \textit{W} states challenging is the second term in the key rate formula (\ref{eq:keyrate}). While the parties can extract perfectly correlated key bits from a noiseless GHZ state, they cannot do it from a \textit{W} state even if the state is ideal. Therefore, error correction is inevitable for multipartite DI-QKD protocols using \textit{W} states. Explicitly, when all the parties perform $\sigma_X$ measurement on a \textit{W} state, the error correction cost is
\begin{equation}
    h \left( \frac{1}{2} - \frac{1}{N} \right),
\end{equation}
where $h$ is the binary entropy and $N$ is the number of parties~\cite{Grasselli2019}.

To realize multipartite DI-QKD protocols that use \textit{W} states, the first term in (\ref{eq:keyrate}) must be sufficiently large to compensate the error correction cost. The first term can be calculated by the violation of Bell inequalities~\cite{Pironio2009, Masanes2011, Tan2021, Brown2021, Brown2024}. To the best of our knowledge, however, there are no known Bell inequalities which make the first term high enough to realize DI-QKD protocols using \textit{W} states. Thus, in this study, we construct Bell inequalities that \textit{W} states can largely violate and make \textit{W}-based DI-QKD protocols possible.

\section{Bell inequalities tailored for \textit{W} states}\label{sec:Bellineq}
\subsection{Constructing Bell expression}
We construct Bell inequalities which give sufficiently large values of the conditional von Neumann entropy in (\ref{eq:keyrate}). First, a Bell inequality $\mathcal{B}$ is generally expressed in the following form
\begin{equation}
    \mathcal{B} = \sum_{x, \boldsymbol{y}, a, \boldsymbol{b}} h_{x\boldsymbol{y}}^{a\boldsymbol{b}} p(a,\boldsymbol{b}|x,\boldsymbol{y}) \leq L_c \leq L_q,
\end{equation}
where $h_{x\boldsymbol{y}}^{a\boldsymbol{b}}$ expresses a coefficient, $L_c$ is a classical limit, $L_q$ is a quantum limit, and we write $\boldsymbol{y} = y_1, \ldots, y_{N-1}$ and $\boldsymbol{b} = b_1, \ldots, b_{N-1}$ for simplicity. We define the following two vectors
\begin{align}
    \boldsymbol{h} &= [h_{x \boldsymbol{y}}^{a \boldsymbol{b}}],\\
    \boldsymbol{p} &= [p(a,\boldsymbol{b}| x, \boldsymbol{y})],
\end{align}
where $\boldsymbol{h}, \boldsymbol{p} \in \mathbb{R}^{m^N d^N}$ and we use the order
\begin{equation}
\begin{split}
    \boldsymbol{p} = [&p(0,\ldots, 0|0,\ldots,0), \, \, ~ \ldots,  p(d,\ldots, d|0,\ldots,0),\\
    &\qquad \qquad \qquad \qquad\quad ~ ~ \vdots\\
    &p(0,\ldots, 0|m,\ldots,m), \ldots, p(d,\ldots, d|m,\ldots,m)
    ].
\end{split}
\end{equation}
Then, we can express a Bell inequality by these vectors
\begin{equation}
    \mathcal{B} = \boldsymbol{h}^T \boldsymbol{p} \leq L_c \leq L_q.
\end{equation}

We search a Bell expression $\boldsymbol{h}$ which gives a Bell inequality largely violated by \textit{W} states. To do so, we take the numerical optimization approach developed in Ref.~\cite{NietoSilleras2014}. In this method, we consider an optimization problem of a guessing probability $p_\text{guess} (A|x^*, E)$, that is, a probability that Eve correctly guesses Alice's measurement outcomes for the key generation rounds. The guessing probability can be used for lower bounding the first term in the key rate formula (\ref{eq:keyrate}) from its relation with the conditional min-entropy~\cite{Konig2009}
\begin{equation}
    \begin{split}
        H(A|x^*, E) &\geq H_\text{min}(A|x^*, E)\\
        &= -\log_2 (p_\text{guess}(A|x^*, E)).
    \end{split}
\end{equation}
Let $B(\mathcal{H})$ be the set of bounded operators on a Hilbert space $\mathcal{H}$. Then, the guessing probability corresponds to a solution of the following optimization problem
\begin{widetext}
\begin{equation}\label{eq:guessing}
    \begin{split}
        p_\text{guess}(A|x^*,E) = \sup \quad & \sum_a
        \text{Tr} [\rho M_{a|x^*} Z_a] \\
        \text{s.t.} \quad &\text{Tr} [\rho M_{a|x} N_{b_1|y_1} \cdots  N_{b_{N-1}|y_{N-1}}] = p(a, \boldsymbol{b}| x, \boldsymbol{y}) \qquad \text{for all} ~ a, b_i, x, y_i\\
        &M_{a|x} \geq 0, \quad N_{b_i|y_i} \geq 0, \quad Z_c \geq 0 \qquad \text{for all} ~ a, b_i, c, x, y_i\\
        &\sum_a M_{a|x} = \sum_{b_i} N_{b_i|y_i} = \sum_c Z_c = I \qquad \text{for all} ~ x, y_i\\
        &\left[M_{a|x}, N_{b_i|y_i} \right] = \left[M_{a|x}, Z_{c} \right] = \left[ N_{b_i|y_i}, Z_{c} \right] =0 \qquad \text{for all} ~ a, b_i, c, x, y_i, i\\
    \end{split}
\end{equation}
\end{widetext}
where the supremum is taken over all quantum states $\rho$ and bounded operators $M_{a|x}$, $N_{b_i|y_i}$ and $Z_{c}$. This optimization problem corresponds to searching Eve's optimal strategy which maximizes the guessing probability under the constraint of the probability distribution of the test rounds. 

We do not have any efficient methods to solve this optimization problem. Thus, we relax this problem into a semidefinite programming (SDP) by using the Navascu{\'{e}}s-Pironio-Ac{\'{i}}n (NPA) hierarchy~\cite{Navascues2008}. By using the NPA hierarchy, we can reformulate this optimization problem into SDP where the optimization variable is a so-called moment matrix (see Appendix~\ref{appendix:NPAhierarchy}). Then, for the relaxed SDP, we can consider its dual problem~\cite{Boyd2004}. We give a brief explanation of the duality of SDP in Appendix~\ref{appendix:SDP}. The dual problem is also SDP and we can efficiently solve the problem. The dual problem of the optimization of the guessing probability corresponds to finding an explicit Bell expression which minimizes the guessing probability~\cite{NietoSilleras2014}. Therefore, by extracting a dual solution, we can find an optimal Bell expression for given probability distribution observed in test rounds.

We search Bell expressions tailored for \textit{W} states for (3, 2, 2), (3, 3, 2), (4, 2, 2) and (5, 2, 2) scenarios. Here, we show the explicit result for the (3, 2, 2) scenario. Let us define POVM $\Pi(\theta) = \cos \theta \sigma_Z + \sin \theta \sigma_X$. We calculate probability distribution $p(a,b_1,b_2|x,y_1,y_2)$ for the quantum state $\ket{W_3}$ and the following measurements
\begin{alignat}{2}
    M_{0|0} &= 0.5 (I+\Pi(\pi/2)), & \, M_{1|0} &= I-M_{0|0},\\
    M_{0|1} &= 0.5 (I+\Pi(0)), & \, M_{1|1} &= I-M_{0|1},\\
    N_{0|0} &= 0.5 (I+\Pi(2\pi/3)), & \, N_{1|0} &= I-N_{0|0},\\
    N_{0|1} &= 0.5 (I+\Pi(\pi/3)), & \, N_{1|1} &= I-N_{0|1}.
\end{alignat}
Here, $\text{Bob}_1$ and $\text{Bob}_2$ perform the same measurements and $I$ is the identity. Then, we find the following Bell expression $\boldsymbol{h}$
\begin{equation}
    \begin{alignedat}{4}
        [ -291&.223, &  159&.065, & 159&.065, & -144&.647,\\
  1146&.316, & -701&.375, & -701&.375, & 1235&.672,\\
   149&.171, & -270&.272, & -180&.496, &  197&.614,\\
    80&.769, &  -56&.324, &  -54&.657, &   78&.917,\\
   149&.171, & -180&.496, &  -270&.272, &  197&.614,\\
    80&.769, & -54&.657, &  -56&.324, &  78&.917,\\
  1288&.666, & ~ -2489&.561, & ~ -2489&.561, &  933&.615,\\
  -128&.785, &  167&.864, &  167&.864, & ~ -133&.405,\\
    37&.170, & -55&.645, &  -55&.645, & -28&.406,\\
    37&.801,&  -49&.508, & -49&.508,&  234&.402,\\
   241&.203, & -299&.824, & ~ -1103&.213, &  253&.583,\\
    35&.804, &  -47&.550, &  58&.961, &  55&.206,\\
   241&.203, & ~ -1103&.213, & -299&.824, & 253&.583,\\
    35&.804, &  58&.961, & -47&.550, &  55&.206,\\
   -23&.943, &   -66&.800,&  -66&.800, &  52&.173,\\
    248&.758, & -75&.853, & -75&.853, &  84&.785].
    \end{alignedat}
\end{equation}
We calculate the first term in (\ref{eq:keyrate}) for this Bell expression and we get the conditional entropy of 0.998. When we assume that every party performs $\sigma_X$ measurement for the key generation rounds, the error correction cost corresponding to the second term in (\ref{eq:keyrate}) is 0.650. Therefore, this Bell expression realizes a multipartite DI-QKD protocol for the $(3, 2, 2)$ scenario. We also show results for the other scenarios in Tab.~\ref{tab:keyrate}. It can be observed that multipartite DI-QKD protocols using \textit{W} states are possible for those scenarios.

\begin{table} 
 \begin{center}
   \caption{Values of the first term, the second term and the key rate in (\ref{eq:keyrate}) for $(3, 2, 2), (3, 3, 2), (4, 2, 2)$ and $(5, 2, 2)$ scenarios.}
   \label{tab:keyrate}
  \begin{tabular}{c|c|c|c} \hline
    Scenario & $H(A|x^*,E)$ & $\max_i H(A|B_i, x^*, y^*_i)$ & $r$   \\ \hline
    (3, 2, 2) & 0.998 & 0.650 & 0.348 \\ \hline
    (3, 3, 2) & 0.999 & 0.650 & 0.349 \\ \hline
    (4, 2, 2) & 0.999 & 0.811 & 0.188 \\ \hline
    (5, 2, 2) & 0.999 &  0.881 & 0.118 \\  \hline
  \end{tabular}
 \end{center}
\end{table}

\subsection{Calculating classical and quantum limits}
The above method gives a Bell expression $\boldsymbol{h}$, but does not give classical and quantum limits of a Bell inequality corresponding to the Bell expression. Here, we calculate those limits of the Bell expressions which we find.

First, we explain how to calculate a classical limit $L_c$. We define the following vector $\boldsymbol{v}_k \in \mathbb{R}^{m^N d^N}$ where $(k \in \{ 1, 2, \ldots, d^{Nm} \})$ which corresponds to extremal points of the polytope of classical correlations~\cite{Brunner2014}. Denote $k = (a_1, \ldots, a_m, b_1^1, \ldots, b_m^1, \ldots, b_1^{N-1}, \ldots, b_m^{N-1})$ where $a_x$ expresses Alice's measurement output when she chooses input $x$ and $b^i_{y_i}$ is $\text{Bob}_i$'s measurement output when he chooses input $y_i$ for $i \in \{1, \ldots, N-1 \}$. Then, entries of $\boldsymbol{v}_k$ can be expressed as follows
\begin{equation}
    v_k (a, \boldsymbol{b}|x, \boldsymbol{y}) = \begin{cases}
        1 & \text{if} \, a = a_x \, \text{and} \, b_i = b_{y_i}^i ~~ \text{for all} ~ i, \\
        0 & \text{otherwise}.
    \end{cases}
\end{equation}
Then, a classical correlation $\mathbf{P}_\text{cl}$ can be expressed as a linear combination of these vectors
\begin{equation}
    \mathbf{P}_{\text{cl}} = \sum_{k=1}^{d^{Nm}} \lambda_k \boldsymbol{v}_k,
\end{equation}
where $\lambda_k \geq 0$ and $\sum \lambda_k = 1$. For a given Bell expression $\boldsymbol{h}^T \boldsymbol{p}$, we can calculate its classical limit by the following linear program
\begin{equation}
    \begin{split}
        \min_{L_c} \quad &L_c\\
        \text{s.t.} \quad &\boldsymbol{h}^T \boldsymbol{v}_k \leq L_c \quad \text{for all} ~ k.
    \end{split}
\end{equation}
A solution of this problem corresponds to a classical limit of the Bell inequality
\begin{equation}
    \mathcal{B} = \boldsymbol{h}^T \boldsymbol{p} \leq L_c.
\end{equation}

Next, we describe how to calculate a quantum limit $L_q$ of a Bell expression. A quantum limit of a Bell expression $\boldsymbol{h}^T \boldsymbol{p} = \sum_{x, \boldsymbol{y}, a, \boldsymbol{b}} h_{x \boldsymbol{y}}^{a \boldsymbol{b}} p(a, \boldsymbol{b}|x, \boldsymbol{y})$ corresponds to a solution of the following problem
\begin{equation}\label{eq:quantumlimit}
    \begin{split}
        \max \quad & \sum_{x, \boldsymbol{y}, a, \boldsymbol{b}} h_{x \boldsymbol{y}}^{a \boldsymbol{b}} \text{Tr} [\rho M_{a|x} N_{b_1|y_1} \cdots N_{b_{N-1}|y_{N-1}}] \\
        \text{s.t.} \quad
        &\sum_a M_{a|x} = \sum_{b_i} N_{b_i|y_i} = I \quad \text{for all} ~ x, y_i\\
        &M_{a|x} \geq 0, \quad N_{b_i|y_i} \geq 0 \quad \text{for all} ~ a, b_i, x, y_i\\
        &\left[M_{a|x}, N_{b_i|y_i} \right] =0 \quad \text{for all} ~ a, b_i, x, y_i\\
        &M_{a|x}, N_{b_i|y_i}\in B(\mathcal{H}) \quad \text{for all} ~ a, b_i, x, y_i
    \end{split}
\end{equation}
where the maximum is taken over all quantum states $\rho$ and operators $M_{a|x}, N_{b_1|y_1}, \ldots, N_{b_{N-1}|y_{N-1}}$. Since the objective function of this optimization problem can be expressed as a linear combination of entries of a moment matrix, we can relax this problem into SDP by using the NPA hierarchy. We show the results of calculating classical and quantum limits of the Bell expressions we find in Tab.~\ref{tab:limit}. Also, we show maximal violations of these Bell expressions with \textit{W} states $L_W$. We can see that, for all scenarios, maximal Bell values with \textit{W} states $L_W$ almost coincide with quantum limits $L_q$. However, $L_q$ is slightly higher than $L_W$ for all scenarios. This can be attributed to a finite NPA hierarchy level. We consider 3rd level NPA hierarchy for these calculations. Therefore, if we consider NPA hierarchy of infinite level, $L_q$ of the Bell expressions might coincide with $L_W$. Analytical analysis of quantum limits of these Bell inequalities is interesting future work.

\begin{table}
 \begin{center}
   \caption{Classical limit $L_c$, violation with \textit{W} states $L_W$ and quantum limit $L_q$ for Bell expressions which we find for $(3, 2, 2), (3, 3, 2), (4, 2, 2)$ and $(5, 2, 2)$ scenarios.}
    \label{tab:limit}
  \begin{tabular}{c|c|c|c} \hline
    Scenario & $L_c$ & $L_W$ & $L_q$   \\ \hline
    (3, 2, 2) & 1791.419 & 1794.3624230 & 1794.3624540 \\ \hline
    (3, 3, 2) & 1713.456 & 1831.4136783 & 1831.4136786 \\ \hline
    (4, 2, 2) & 1476.978 & 1514.9137399 & 1514.9137408 \\ \hline
    (5, 2, 2) & 2409.119 & 2441.8495037 & 2441.8495040 \\ \hline
  \end{tabular}
 \end{center}
\end{table}

\section{Analyzing required detection efficiency}\label{sec:required}
In this section, we consider realistic detectors with imperfect detection efficiency and analyze the minimum required detection efficiency for multipartite DI-QKD protocols using \textit{W} states. 
The guessing probability approach used in the previous section works pretty well if there is no experimental imperfections. 
However, we find that this approach is highly susceptible to imperfections including finite detection efficiency.
Therefore, in this section, we take another approach to calculate key rates, that is, the numerical optimization based on the quasi-relative entropy developed in Ref.~\cite{Brown2024}. 
Here, we consider that an $N$-partite \textit{W} state is distributed by the direct transmission where the \textit{W} state is locally generated at some place and distributed among the parties (Fig.~\ref{fig:schematic}(a)). We assume unit success probability, $P_\text{succ} = 1$, for the \textit{W} state generation and the state is shared by each party without any transmission losses and noises (i.e. $\eta=1$ in Fig.~\ref{fig:schematic}(a)). 
We only consider the non-unit detection efficiency as the imperfection. 
%We can use the guessing probability to calculate key rates. While the guessing probability works pretty well when we do not consider any experimental imperfections, however, we find that it is highly susceptible to losses such as detection efficiency. Therefore, in this section, we take another approach to calculate key rates, that is, the numerical optimization based on the quasi-relative entropy developed in Ref.~\cite{Brown2024}. Here, we consider that an $N$-partite \textit{W} state is distributed by the direct transmission, that is, the \textit{W} state is locally generated and distributed among the parties. We assume that we can locally generate a \textit{W} state with unit probability $P_\text{succ} = 1$.

\subsection{Measurement modeling}
First, let us explain how to model detection efficiency. We consider scenarios where each party has two measurment settings with two outcomes, i.e., $(N, 2, 2)$ scenarios. For $i \in \{1, \ldots, N-1 \}$, we define POVMs corresponding to Alice's and $\text{Bob}_i$'s measurements as follows
\begin{alignat}{2}\label{eq:arbitrarymeasurement}
    M_{0|x} &= 0.5 \eta_e (I+\Pi(\theta_x^A)), & \, M_{1|x} &= I-M_{0|x},\\
    N_{0|y_i} &= 0.5 \eta_e (I+\Pi(\theta_{y_i}^{B_i})), & \, N_{1|y_i} &= I-N_{0|y_i},
\end{alignat}
where $\eta_e$ corresponds to detection efficiency and $\Pi(\theta) = \cos \theta \sigma_Z + \sin \theta \sigma_X$. We numerically optimize the parameters $\theta_x^A, \theta_{y_1}^{B_1}, \ldots,\theta_{y_{N-1}}^{B_{N-1}}$ to maximize key rates. 
 In addition, we consider another measurement model to analyze experimental feasibility of multipartite DI-QKD protocols using \textit{W} states. One of the practical way of generating \textit{W} states is splitting a single-photon with several beamsplitters. Since \textit{W} states generated in this way are encoded in the photon-number basis, its measurements should also be in the same basis. Although, in principle it is possible to implement arbitrary Pauli measurements in such basis~\cite{Takeoka2006} and its proof-of-principle was demonstrated~\cite{Izumi2020}, its implementation with near-perfect efficiency is still not straightforward. 
Therefore, we consider its approximate version where the measurement consists of optical displacement operation followed by a photon detection. This measurement is expressed as
\begin{equation}
    \{ \ketbra{\alpha}{\alpha} , I-\ketbra{\alpha}{\alpha}  \},
\end{equation}
where $\ket{\alpha}$ is a coherent state
\begin{equation}
    \ket{\alpha} = e^{-\frac{|\alpha|^2}{2}} \sum_{n=0}^{\infty} \frac{\alpha^n}{\sqrt{n!}} \ket{n}.
\end{equation}
Although Bell-inequality violation with this measurement and \textit{W} states was analyzed in Ref.~\cite{Bjerrum2023}, the authors did not consider multipartite DI-QKD protocols. Here, we model the detection efficiency $\eta_e$ as a pure-loss channel with transmissivity $\eta_e$ prior to the displacement operation since we can compensate any losses at a photon detector by increasing the amplitude. After transmitting pure-loss channels with transmissivity $\eta_e$, an $N$-partite \textit{W} state $\ket{W_N}$ comes to be
\begin{equation}
    \ketbra{W_N}{W_N} \rightarrow \eta_e \ketbra{W_N}{W_N} + (1-\eta_e) \ketbra{0_N}{0_N}.
\end{equation}
$\ket{0_N}$ expresses a quantum state where all of the $N$ modes are vacuum states.

\subsection{Key rate calculation}
In this method, we can obtain a lower bound on the conditional von Neumann entropy in (\ref{eq:keyrate}) by solving the following optimization problem~\cite{Brown2024}
\begin{widetext}
\begin{equation}\label{eq:quasirelative}
    \begin{split}
        c_m + \inf \quad &\sum_{i=1}^{m-1} \frac{w_i}{t_i \ln 2 } \sum_{a=0}^1 \text{Tr} [ \rho \hat{M}_{a|x^*} (Z_{a,i} + Z_{a,i}^* +(1-t_i) Z_{a,i}^* Z_{a,i}) + t_i Z_{a,i} Z_{a,i}^* ]\\
        \text{s.t.} \quad
        &\text{Tr} [ \rho M_{a|x} N_{b_1|y_1} \cdots  N_{b_{N-1}|y_{N-1}} ]= p(a, \boldsymbol{b}| x, \boldsymbol{y}) \qquad \text{for all} ~ a, b_i, x, y_i\\
        &\sum_a M_{a|x} = \sum_{b_i} N_{b_i|y_i} = I \qquad \text{for all} ~ x, y_i\\
        &M_{a|x} \geq 0, \quad N_{b_i|y_i} \geq 0 \qquad \text{for all} ~ a, b_i, x, y_i\\
        & Z_{c,i}^* Z_{c,i} \leq \alpha_i^2, \quad Z_{c,i} Z_{c,i}^* \leq \alpha_i^2 \qquad \text{for all} ~ c, i\\
        &\left[M_{a|x}, N_{b_i|y_i} \right] = \left[M_{a|x}, Z_{c,i}^{(*)} \right] = \left[ N_{b_i|y_i}, Z_{c,i}^{(*)} \right] =0 \qquad \text{for all} ~ a, b_i, c, x, y_i, i\\
        &M_{a|x}, N_{b_i|y_i}, Z_{c,i} \in B(\mathcal{H}) \qquad \text{for all} ~ a, b_i, c, x, y_i, i
    \end{split}
\end{equation}
\end{widetext}
where the infimum is taken over all quantum states $\rho$ and bounded operators $M_{a|x}$, $N_{b_i|y_i}$ and $Z_{c,i}$, $t_1, \ldots, t_m$ and $w_1, \ldots, w_m$ are the nodes and weights of an $m$-point Gauss-Radau quadrature on $[0, 1]$ with an end point $t_m = 1$, $\hat{M}_{a|x^*} = (1-p_n) M_{a|x^*} + p_n M_{a \oplus 1 |x^*}$, $c_m = \frac{2p_n (1-p_n)}{m^2 \ln2} + \sum_{i=1}^{m-1} \frac{w_i}{t_i \ln 2}$, $\alpha_i = \frac{3}{2} \max \{ \frac{1}{t_i} , \frac{1}{1-t_i}  \}$ and $p(a, \boldsymbol{b} | x, \boldsymbol{y})$ is an observed probability distribution for the test rounds. This optimization problem is also relaxed into SDP by using the NPA hierarchy.

\subsection{Results}
We show the results for analyzing required detection efficiency. Here, we fix the number of nodes of the Gauss-Radau quadrature $m=8$ and numerically optimize measurements and the probability that Alice flips her key bits $p_n$ to maximize key rates.

\begin{figure}[htbp]
 \centering
 \includegraphics[keepaspectratio, scale=0.5]{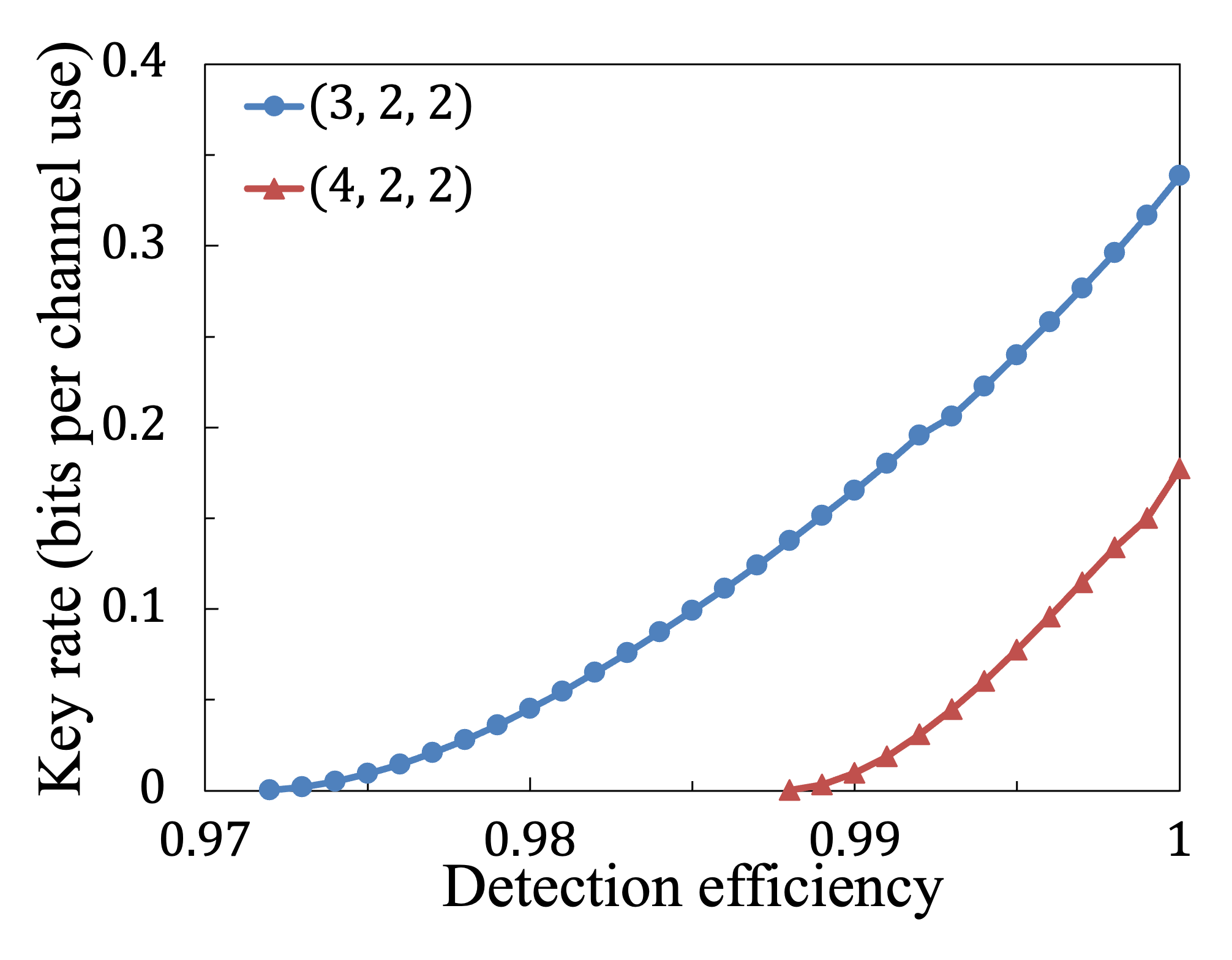}
 \caption{Key rate versus detection efficiency $\eta_e$ for the scenarios (3, 2, 2) and (4, 2, 2) where the legitimate parties perform arbitrary Pauli measurements. Blue circles show key rates of the (3, 2, 2) scenario and red triangles show key rates of the (4, 2, 2) scenario. We numerically optimize the measurement parameters $\theta_x^A, \theta_{y_1}^{B_1}, \ldots, \theta_{y_{N-1}}^{B_{N-1}}$ and the probability $p_n$ to maximize key rates.}
 \label{fig:DirectPauli}
\end{figure}

First, we show the results where every party can perform arbitrary Pauli measurements on \textit{W} states in Fig.~\ref{fig:DirectPauli}. Blue circles show key rates of the (3, 2, 2) scenario and red triangles show key rates of the (4, 2, 2) scenario. We find that the minimum required detection efficiencies to generate a secret key are 97.2 \% and 98.8 \% for the settings $(3, 2, 2)$ and $(4, 2, 2)$, respectively. The optimal measurement for the key generation rounds is close to Pauli $\sigma_X$ measurement. The higher detection efficiency is required for the higher $N$. These can be interpreted as follows. If we consider only the error correction cost and the number of parties $N$ is sufficiently large, an optimal measurement for the key generation rounds is Pauli $\sigma_Z$ measurement since almost all parties can obtain the same measurement outcome $``0"$. However, this means that an eavesdropper easily guesses Alice's measurement outcomes, making the conditional entropy $H(A|x^*, E)$ small. Therefore, Pauli $\sigma_X$ measurement is optimal if we consider the net effect on key rates since we get large values of the conditional von Neumann entropy. When every party performs the $\sigma_X$ measurement for the key generation rounds, the error correction cost is $h ( 1/2 -1 /N )$ and this increases with the number of parties. Then, we require higher detection efficiency for larger number of parties. While achieving detection efficiencies over 97 \% is quite challenging with fully optical setups, we can realize such high detection efficiencies with matter-based systems~\cite{Nadlinger2022, Zhang2022, Lu2026}.

\begin{figure}[htbp]
 \centering
 \includegraphics[keepaspectratio, scale=0.5]{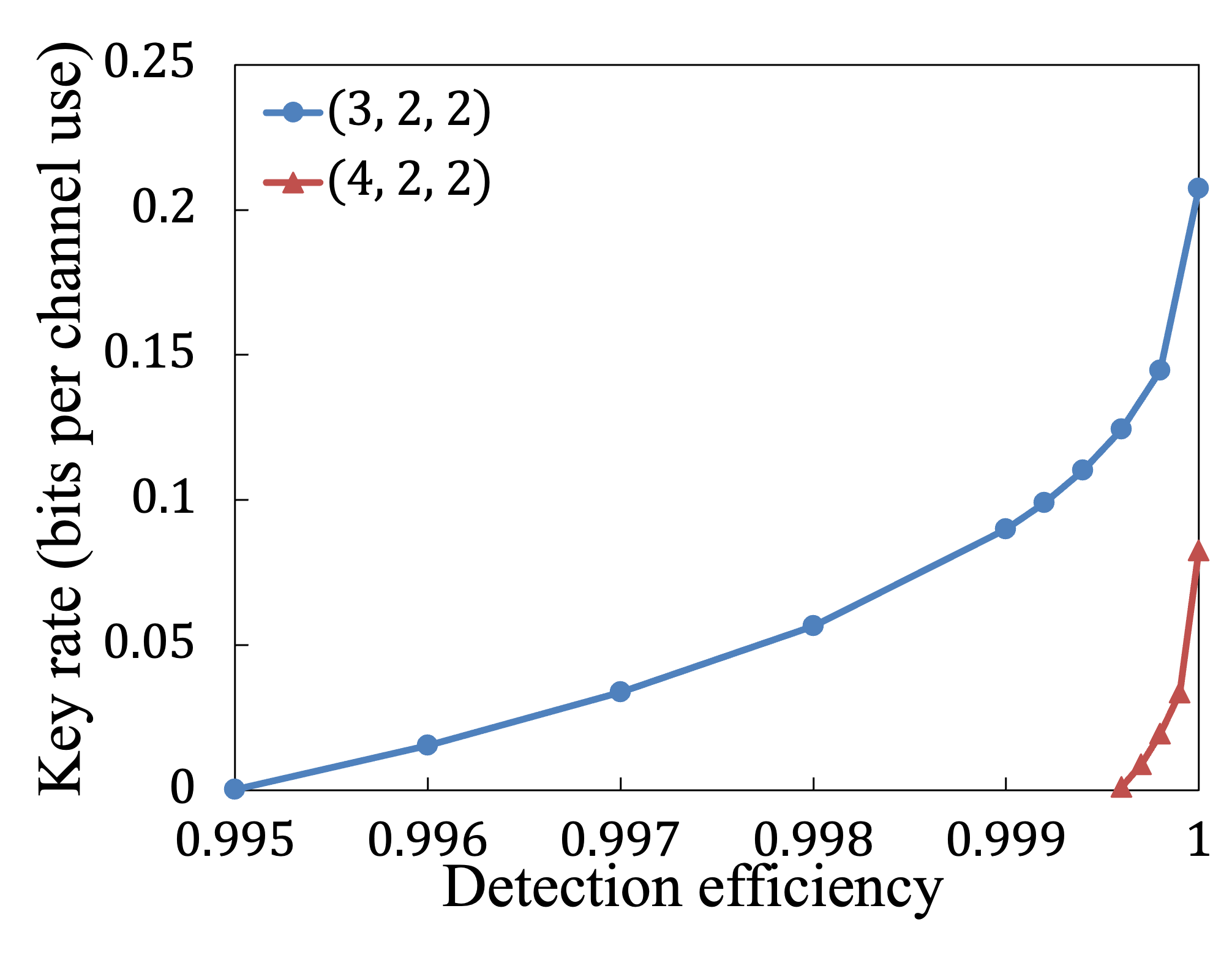}
 \caption{Key rate versus detection efficiency $\eta_e$ for the scenarios (3, 2, 2) and (4, 2, 2) where the legitimate parties perform the displacement-based measurements. Blue circles show key rates of the (3, 2, 2) scenario and red triangles show key rates of the (4, 2, 2) scenario. We numerically optimize the displacements $\alpha$ of all the legitimate parties and the probability $p_n$ to maximize key rates.}
 \label{fig:DirectDisplacement}
\end{figure}

In Fig.~\ref{fig:DirectDisplacement}, we show key rates for $(3, 2, 2)$ and $(4, 2, 2)$ scenarios where the remote parties perform the displacement-based measurements. Blue circles and red triangles show key rates of the scenarios $(3, 2, 2)$ and $(4, 2, 2)$, respectively. We find minimum required detection efficiency is 99.5 \% for the $(3, 2, 2)$ scenario and 99.96 \% for the $(4, 2, 2)$ scenario. These values are higher than those of the situation where the parties can perform arbitrary Pauli measurements since we cannot perform such arbitrary measurements with displacement operations and photon detectors.

% \begin{figure}[htbp]
%  \centering
%  \includegraphics[keepaspectratio, scale=0.5]{figure/DICKAwithW/DirectTransmission.png}
%  \caption{Direct transmission. $p_{DC} = 10^{-6}$}
%  \label{fig:DirectTransmission}
% \end{figure}

\section{Long-distance multipartite DI-QKD with \textit{W} states}\label{sec:roga}
So far, we consider that $N$ parties share a \textit{W} state by the direct transmission, that is, the situation where an $N$-partite \textit{W} state is locally generated at a central station and distributed among the parties. However, if channel loss between the central station and each party is $\eta$, distribution rate of the \textit{W} state scales as $O( \eta^N)$, decreasing exponentially with the number of parties. Recently, we propose the efficient distribution scheme of \textit{W} states in Ref.~\cite{Roga2023}. We call this RIHT protocol. The parties can share an $N$-partite \textit{W} state with distribution rate $O(\eta)$ in this protocol. Therefore, in this section, we consider a multipartite DI-QKD protocol with the RIHT protocol.

\begin{figure}[htbp]
 \centering
 \includegraphics[keepaspectratio, scale=0.4]{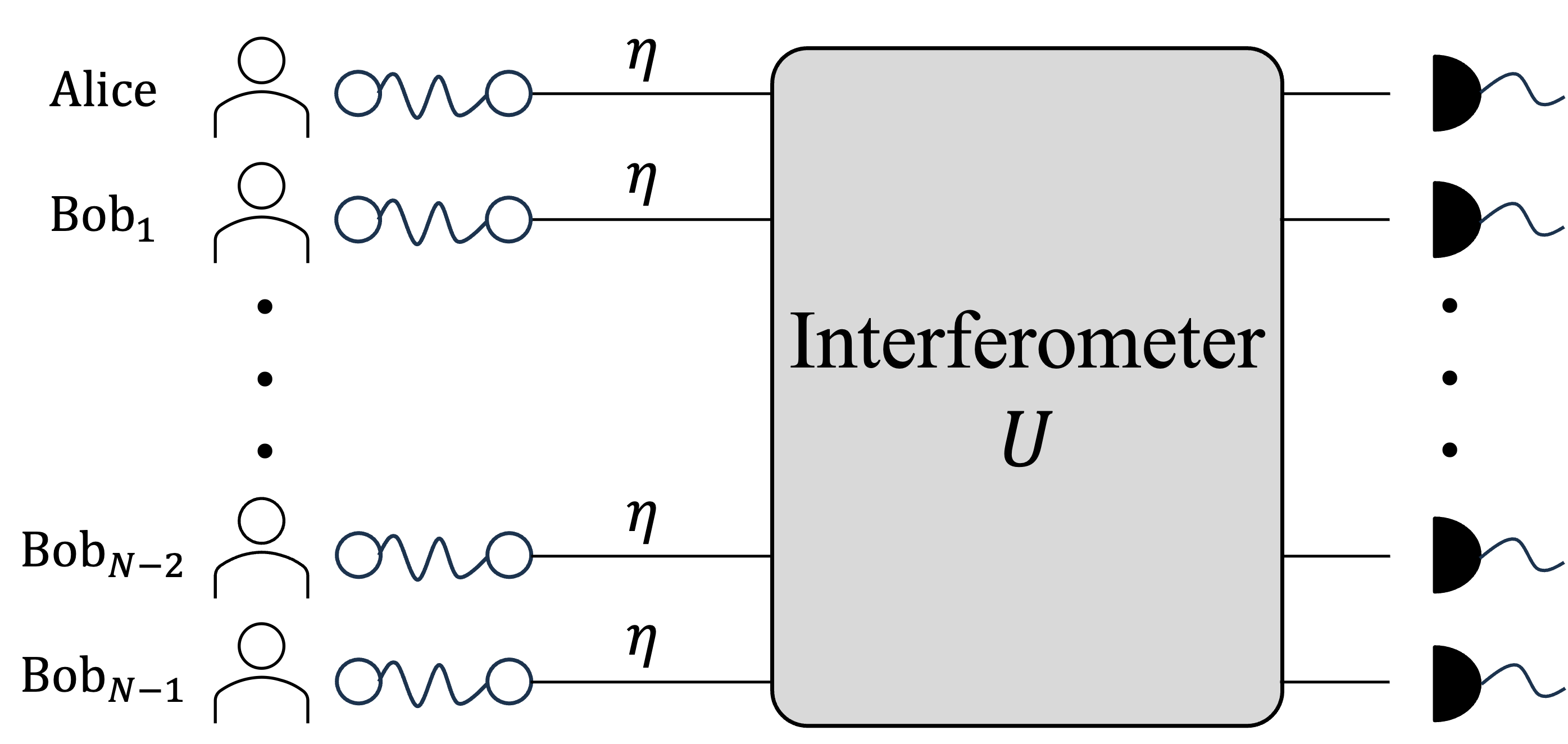}
 \caption{Schematic of the RIHT protocol. Every party prepares a two-mode entangled state $\sqrt{q} \ket{00} + \sqrt{1-q} \ket{11}$ and transmits one part to a central station through a pure-loss channel with transmissivity $\eta$. The central station is composed of an interferometer $U$ with 50:50 beamsplitters and $N$ single-photon detectors. The parties distil a secret key from events where only one of the single-photon detectors detects a single photon.}
 \label{fig:DICKAwithWsetup}
\end{figure}

A schematic of the RIHT protocol is shown in Fig.~\ref{fig:DICKAwithWsetup}. Same as the direct transmission, we consider a $(N, m, d)$ scenario where $N$ parties share a common secret key and each party has $m$ different measurement settings with $d$ outcomes. First, each party prepares an entangled state of two modes, $X$ and $X'$,
\begin{equation}\label{eq:entanglement}
    \sqrt{q} \ket{00}_{XX'} + \sqrt{1-q} \ket{11}_{XX'},
\end{equation}
where $q \approx1$ is a parameter we can choose freely, mode $X$ is kept at each party's site and mode $X'$ is sent to a central station. Then, each party transmits mode $X'$ to the central station through a pure-loss channel with transmissivity $\eta$. The central station is composed of an interferometer $U$ with 50:50 beamsplitters and $N$ single-photon detectors with dark count probability $p_d$. When $N=2^n$, an operation of the interferometer $U$ can be expressed as
\begin{equation}
    U = u^{\otimes n},
\end{equation}
where
\begin{equation}
    \begin{split}
        u = \frac{1}{\sqrt{2}} \left( \begin{array}{cc}
            1 & 1 \\
            -1 & 1
        \end{array} \right).
    \end{split}
\end{equation}
When one of the $N$ single-photon detectors detects a single photon, the $N$-partite \textit{W} state $\ket{W_N}$ is distributed among the parties. We call such events successful events. This is a multipartite extension of the single-photon interference which is widely used in QKD protocols such as twin-field QKD~\cite{Lucamarini2018}.  When successful events occur, each party performs measurements on the distributed \textit{W} state in the same way as the direct transmission. Moreover, Alice flips her bits for the key generation rounds with probability $p_n$. Finally, the parties perform error correction and privacy amplification, and distil a common secret key as we do it for the direct transmission scenario. We calculate an explicit expression of a distributed \textit{W} state for $N=3, 4$ in Appendix~\ref{appendix:statecalculation}.

We can calculate key rates of this protocol same as the direct transmission. We consider all the parties perform arbitrary Pauli measurements expressed as (\ref{eq:arbitrarymeasurement}) with unit efficiency $\eta_e = 1$. In an experiment, we can realize such measurements by using matter qubits. We numerically optimize the measurement parameters of the parties and the probability $p_n$ to maximize key rates.

\subsection{Results}
\begin{figure}[htbp]
 \centering
 \includegraphics[keepaspectratio, scale=0.5]{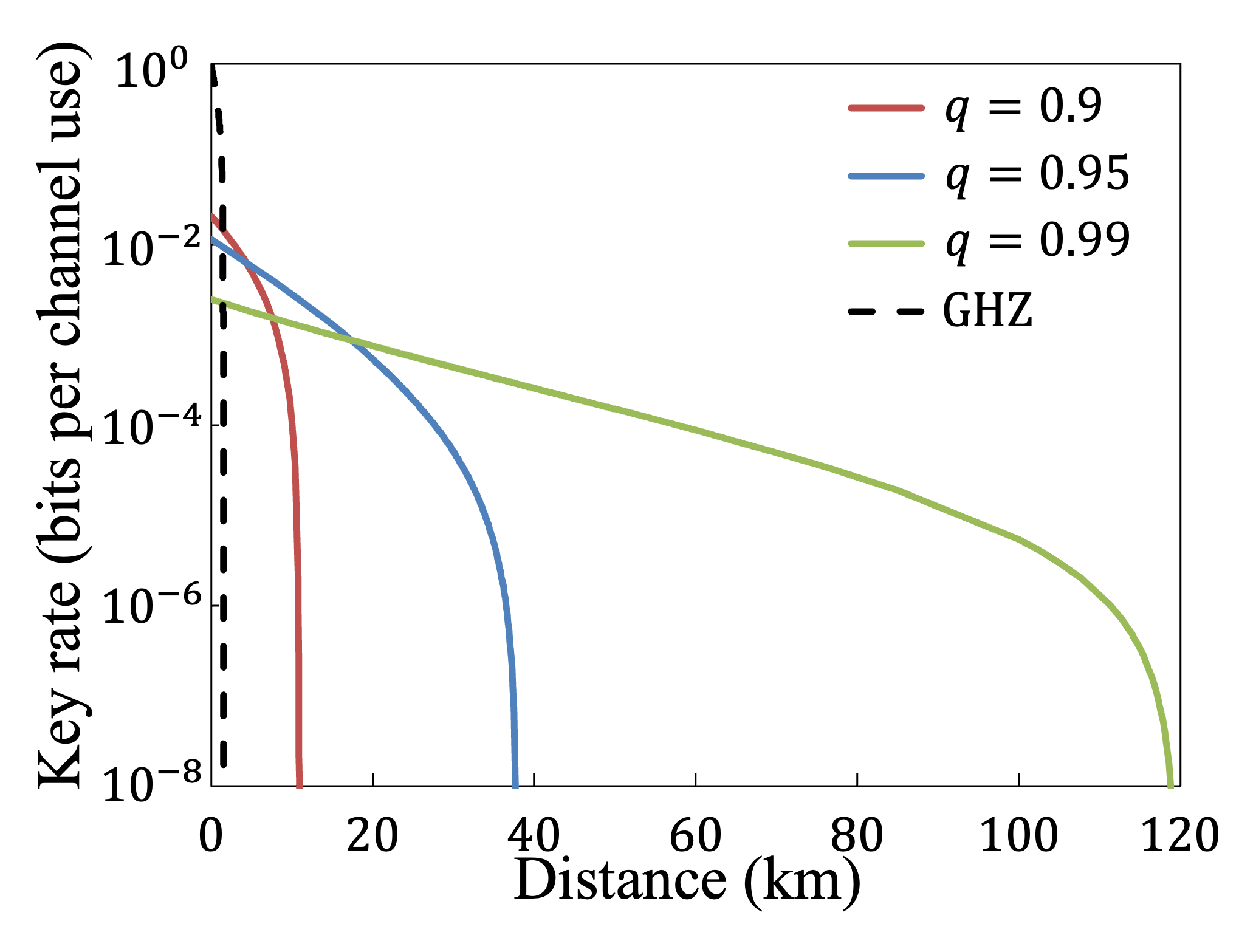}
 \caption{Key rate with the RIHT protocol versus distance $L$ for the scenario (3, 2, 2) for different values of the parameter $q$ of the entangled state which each party prepares. We use the detection efficiency $\eta_e = 1$ and the dark count probability $p_d = 10^{-6}$. Black dashed line shows key rates of a multipartite DI-QKD protocol with locally generated GHZ states.}
 \label{fig:Roga3}
\end{figure}

\begin{figure}[htbp]
 \centering
 \includegraphics[keepaspectratio, scale=0.5]{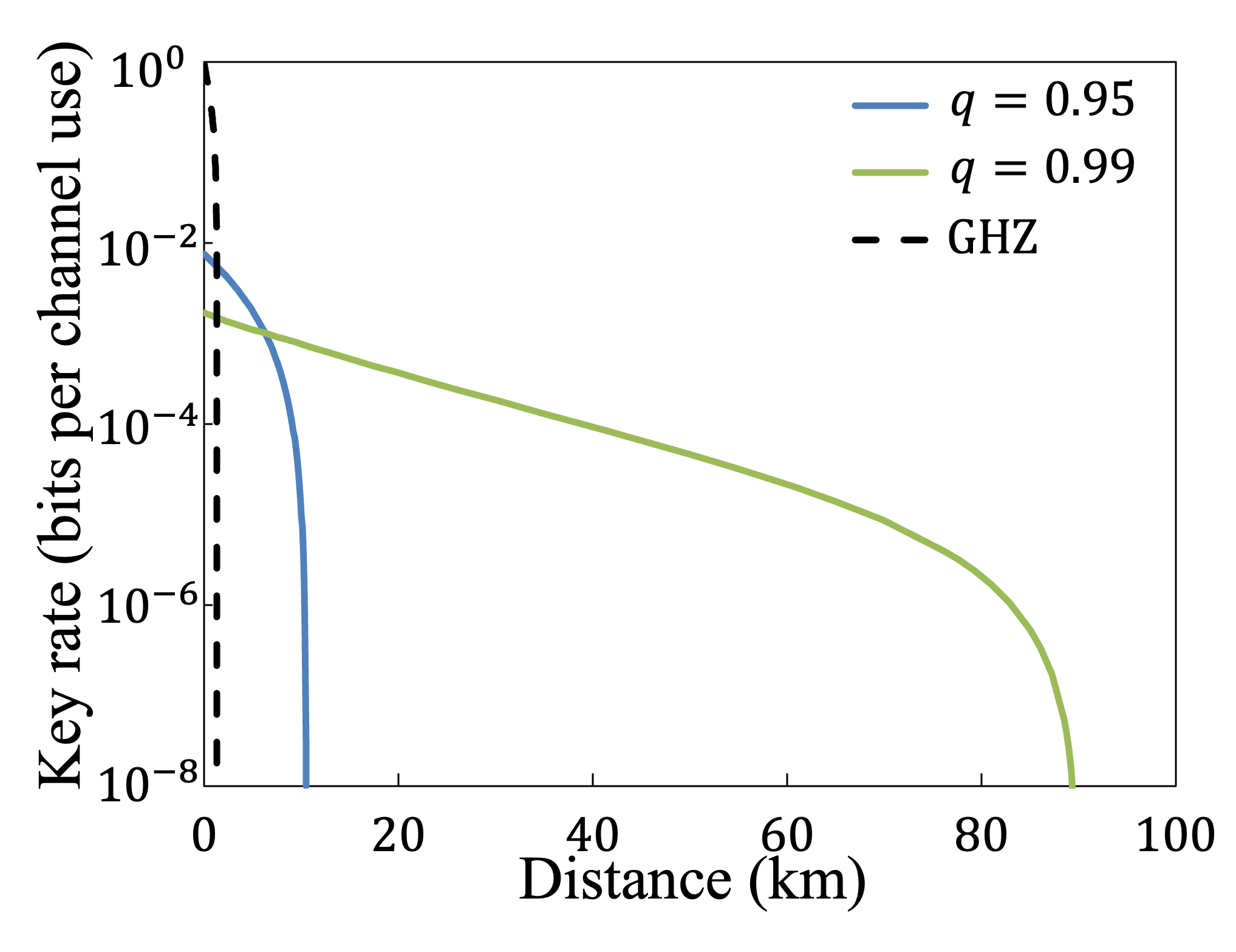}
 \caption{Key rate with the RIHT protocol versus distance $L$ for the scenario (4, 2, 2) for different values of the parameter $q$ of the entangled state which each party prepares. We use the detection efficiency $\eta_e = 1$ and the dark count probability $p_d = 10^{-6}$. Black dashed line shows key rates of a multipartite DI-QKD protocol with locally generated GHZ states.}
 \label{fig:Roga4}
\end{figure}

In Fig.~\ref{fig:Roga3} and Fig.~\ref{fig:Roga4}, we show key rates of the (3, 2, 2) scenario and the (4, 2, 2) scenario, respectively, for different values of the parameter $q$ in (\ref{eq:entanglement}). We plot key rates against the distance between each party and the central station $L$ in kilometers which relates with the transmittance $\eta$ as $\eta = 10^{-0.02 L}$ corresponding to a standard optical fiber with loss of 0.2 dB/km. We fix the dark count probability of the single-photon detectors at the central station $p_d = 10^{-6}$. We also plot key rates of a multipartite DI-QKD protocol with GHZ states where a GHZ state is locally generated and distributed among legitimate parties, and key rates are calculated from violations of a parity-CHSH inequality~\cite{Ribeiro2018, Holz2019, Ribeiro2019}. It can be observed that the key rates with the RIHT protocol outperform the GHZ-based protocol. While the GHZ protocol distributes a secret key over few kilometers, our protocol achieves the range longer than 100 kilometers. This is because that we can effectively consider the effect of channel losses as probabilistic generation of \textit{W} states by using the single-photon interference. Also, we find that the key rates significantly change depending on the parameter $q$. The legitimate parties can share a \textit{W} state with high fidelity when $q$ is small since small $q$ means that the probability that each party sends a single photon to the central station is small. Fidelity of a distributed \textit{W} state decreases because of multiphoton events, i.e., more than one party send photons and only one photon arrives at the central station. Although these events are classified into the successful events, distributed states are different from the desired \textit{W} states. When $q$ is large, the effect of such multiphoton events is large and we cannot distribute a secret key over long distances. However, large $q$ means that we can obtain large success probability and key rates at short distances are large compared to key rates where $q$ is small. Same as the direct transmission protocol, we get smaller key rates as the number of parties $N$ increases.

\section{Long-distance DI-QKD using \textit{W} states with Gaussian formalism}\label{sec:gaussian}
In this section, we investigate feasibility of the RIHT protocol in a fully optical setup. 
In a fully optical setup, the SPDC process is widely used, which generates entangled state called the two-mode squeezed vacuum (TMSV). 
TMSV is in a class of Gaussian states and we summarize the Gaussian formalism and the detailed analyses in this section in Appendix~\ref{appendix:Gaussian}.
Here, we consider a protocol which is equivalent to the one in the previous section except that each party prepares the TMSV with mean photon number $\bar{n}$ instead of the qubit entangled state in (\ref{eq:entanglement}) and performs displacement-based measurements instead of arbitrary Pauli measurements. Also, the detectors at the central station are now on-off detectors which only distinguish zero or one or more photons.

Since the operation of a beamsplitter is a Gaussian operation, a quantum state after the channel transmission and the operation of the interferometer $U$ is a Gaussian state, that is, completely characterized by its displacement vector and covariance matrix. Measurement operators of an on-off detector is expressed as 
\begin{equation}
    \{ (1-p_d) \ketbra{0}{0} , I - (1-p_d) \ketbra{0}{0}  \},
\end{equation}
where $p_d$ represents the dark-count probability. Then, by denoting the quantum state before considering detection at the central station $\rho_{\boldsymbol{X} \boldsymbol{X'}}$ where $\boldsymbol{X}$ expresses modes which are measured by the parties and $\boldsymbol{X'}$ expresses modes which are sent to the central station, we can express an unnormalized quantum state after the detection as follows
\begin{equation}
\begin{split}
    \tilde{\sigma}_{\boldsymbol{X}} &= \text{Tr} [\rho_{\boldsymbol{X} \boldsymbol{X'} } \{  (1-p_d)^{N-1} \ketbra{0\cdots 0}{0 \cdots 0}_{\boldsymbol{X'}/D_1} \\
    &\quad -(1-p_d)^{N} \ketbra{0 \cdots 0}{0 \cdots 0}_{\boldsymbol{X'}} \}],
\end{split}
\end{equation}
where $\ket{0}$ means a vacuum state and $\boldsymbol{X'}/D_1$ means all modes except for the mode $D_1$ where the on-off detector clicks. The success probability $P_\text{succ}$ is expressed as a trace of this quantum state
\begin{equation}
    P_\text{succ} = \text{Tr} [\tilde{\sigma}_{\boldsymbol{X}}].
\end{equation}
From a quantum state $\sigma$ after we normalize $\tilde{\sigma}$, we can calculate a probability distribution $p(a, \boldsymbol{b}|x, \boldsymbol{y})$. By using the probability distribution, we can calculate key rates for this protocol. For an explicit calculation of the probability distribution of this protocol, see Appendix~\ref{appendix:Gaussian}.

In Fig.~\ref{fig:Roga3Gaussian}, we show results of key rate calculation for this protocol. We consider the $(3, 2, 2)$ scenario and the mean photon number of two-mode squeezed states $\bar{n} = 0.01$. Blue solid line shows key rates of the RIHT protocol with Gaussian states and measurements, and black dashed line shows key rates of the GHZ-based protocol where the ideal GHZ states are distributed by direct transmissions. We find that the RIHT protocol achieves longer distances than that of the ideal-GHZ-based protocol even with experimentally feasible states and measurements.

\begin{figure}[htbp]
 \centering
 \includegraphics[keepaspectratio, scale=0.5]{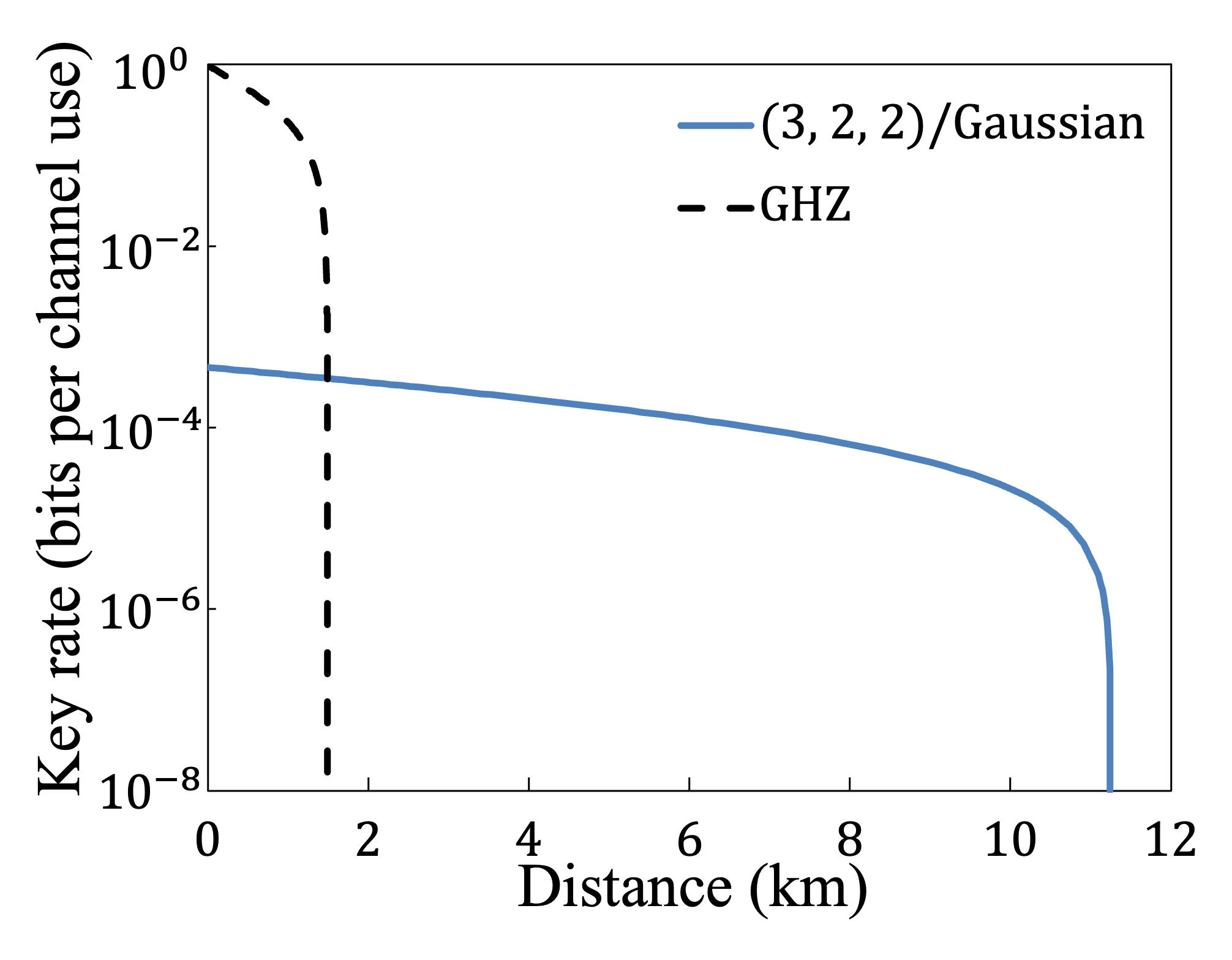}
 \caption{Key rate with the RIHT protocol versus distance $L$ for the setting (3, 2, 2) with Gaussian states and measurements. We use the detection efficiency $\eta_e = 1$ and the dark count probability $p_d = 10^{-6}$. Black dashed line shows key rates of a multipartite DI-QKD protocol with GHZ states.}
 \label{fig:Roga3Gaussian}
\end{figure}

\section{Conclusion}\label{sec:conclusion}
In this paper, we theoretically demonstrate that one can construct multipartite DI-QKD with {\it W} states. 
We numerically construct Bell inequalities which \textit{W} states violate largely. By using these Bell inequalities, we construct a multipartite DI-QKD protocol and show that one can extract device-independent secret keys from \textit{W} states. We also discuss the realistic situations. First, we analyze the required detection efficiencies for {\it W}-based DI-QKD protocols when {\it W} states are ideally distributed. In addition, we take into account the channel losses for {\it W} state distribution. We propose to use the RIHT protocol for efficient distribution of {\it W} states over lossy channels and show that this protocol achieves longer distances than a multipartite DI-QKD protocol based on GHZ states even with experimentally feasible implementations. Our results open a potential of {\it W} states in DI-QKD.i

There are several interesting perspectives of this research. 
In this paper, we numerically show that the quantum limits of our Bell inequalities are almost saturated by {\it W} states. 
From the fundamental viewpoint, an important direction is to prove this analytically. It will give us better understanding of the relationship between \textit{W} states and Bell inequalities. Improvement of the {\it W} state-based multipartite DI-QKD or application of the other multipartite entangled states such as Dicke states and $N00N$ states are also interesting future work in practical side.

\begin{acknowledgments}
This work is supported by JST SPRING, Grant No. JPMJSP2123, JST Moonshot R\&D, Grant No. JPMJMS226C and Grant No. JPMJMS2061, JST CRONOS, Grant No. JPMJCS24N6, JST ASPIRE, Grant No. JPMJAP2427, and JST COI-NEXT, Grant No. JPMJPF2221.
\end{acknowledgments}

\appendix

\section{NPA hierarchy}\label{appendix:NPAhierarchy}
Let us explain the NPA hierarchy. In this paper, we use the NPA hierarchy to calculate the conditional von Neumann entropy in (\ref{eq:keyrate}) and quantum limits of Bell inequalities. Full explanation can be found in the original paper~\cite{Navascues2008}.

First, we construct an optimization problem which is relaxed into SDP later by using the NPA hierarchy. Let $\mathcal{H}$ denote a Hilbert space and $\rho$ be a quantum state on $\mathcal{H}$. We define a set of operators on the same Hilbert space $\mathcal{O}$. Also, we consider expectation values of operators in this set $\ev{O_i^\dagger O_j} = \text{Tr} [\rho O_i^\dagger O_j]$ where $O_i \in \mathcal{O}$. We consider an optimization problem where we minimize a linear combination of these expectation values over all quantum states $\rho$ and operators $O_i$. The objective functions of the optimization problems in (\ref{eq:guessing}), (\ref{eq:quantumlimit}), and (\ref{eq:quasirelative}) are expressed in this way. Since we optimize over all quantum states and operators without any constraints on the dimension of the Hilbert space, this optimization is not feasible. Then, by using the NPA hierarchy, we relax this optimization problem into SDP which is efficiently solved. In the $k$-th level NPA hierarchy, we choose $l$ operators from a set of operators which is composed of Alice's measurement operators, Bobs' measurement operators, Eve's measurement operators and the identity operator for $1 \leq l \leq k$. Then, we consider a set of products of $l$ those operators $\mathcal{O}_k$. We construct a $k$-th moment matrix $\Gamma^{(k)}$ whose entries are $\Gamma_{ij}^{(k)} = \ev{(O_i^{(k)})^\dagger O_j^{(k)}}$ where $O_i^{(k)} \in \mathcal{O}_k$. It is known that a moment matrix constructed in this way is positive semidefinite. Also, the objective function of the optimization problem is expressed as a linear combination of entries of a moment matrix. Therefore, the optimization problem is considered as SDP whose variable is a moment matrix. Here, note that a $k$-th level moment matrix does not necessarily correspond to feasible points of the original optimization problem and the range of optimization in the $k$-th level NPA hierarchy is wider than that of the original optimization problem. In the limit of $k \to \infty$, however, it is known that optimal values of these optimization problems coincide. We perform the NPA hierarchy by using NCPOL2SDPA~\cite{Wittek2015} and solve SDPs by using MOSEK~\cite{Mosek2024}.

\section{Duality of semidefinite programming}\label{appendix:SDP}
We describe the duality of SDP~\cite{Boyd2004}. We consider the following SDP
\begin{equation}
\begin{split}
    \min_{X} \quad &\text{Tr} [C X] \\
    \text{s.t.} \quad &\text{Tr} [A_i X] = b_i \quad (i = 1, \ldots, m)\\
    & X \geq 0
\end{split}
\end{equation}
where $C$ and $A_i$ are $n \times n$ matrices, $b \in \mathbb{R}^m$ is a constraint vector, and we minimize over all positive semidefinite matrices $X$. We call this optimization a primal problem. We can construct a dual problem corresponding to the primal problem as follows
\begin{equation}
\begin{split}
    \max_{y, Z} \quad &\sum_{i=1}^m y_i b_i \\
    \text{s.t.} \quad &C - \sum_i y_i A_i - Z = 0\\
    & Z \geq 0
\end{split}
\end{equation}
where we maximize over all vectors $y$ and positive semidefinite matrices $Z$.

There is an important relationship between the primal and dual problems, that is, the following inequality holds
\begin{equation}
    \text{Tr} [CX] \geq \sum_{i=1}^m y_i b_i.
\end{equation}
In other words, an optimal value of the primal problem is not smaller than that of the dual problem. This inequality is derived in the following way.
\begin{equation}
    \begin{split}
        \text{Tr} [CX] - \sum_i b_i y_i &= \text{Tr} [ (\sum_i y_i A_i + Z)X] - \sum_i b_i y_i \\
        &= \sum_i b_i y_i + \text{Tr} [ZX] - \sum_i b_iy_i \\
        &= \text{Tr} [ZX]\\
        &\geq 0
    \end{split}
\end{equation}
Here, the inequality holds since $\text{Tr} [AB] \geq 0$ when both $A$ and $B$ are positive semidefinite.

\section{State calculation for RIHT protocol}\label{appendix:statecalculation}

\begin{figure}[htbp]
 \centering
 \includegraphics[keepaspectratio, scale=0.45]{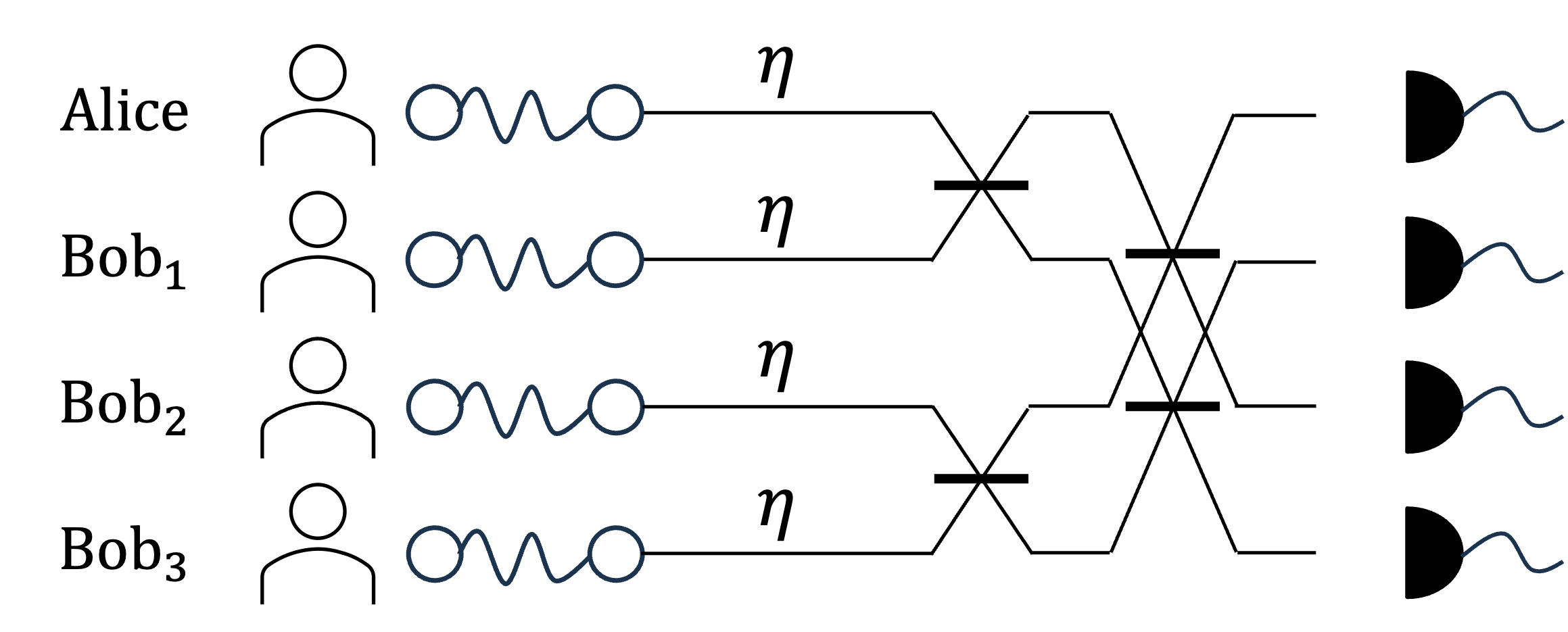}
 \caption{Schematic of RIHT protocol where $N=4$. Every party prepares a two-mode entangled state $\sqrt{q} \ket{00} + \sqrt{1-q} \ket{11}$ and transmits one part to a central station through a pure-loss channel with transmissivity $\eta$. The central station is composed of an interferometer with four 50:50 beamsplitters and four single-photon detectors. The parties distil a secret key from events where only one of the single-photon detectors detects a single photon.}
 \label{fig:DICKAwithWsetup4}
\end{figure}

We calculate a joint quantum state that the parties share when successful events occur in the RIHT protocol. Here, we provide an argument for $N=4$ shown in Fig. \ref{fig:DICKAwithWsetup4}. 

First, each party prepares an entangled state in the following form
\begin{equation}
    \sqrt{q}\ket{00}_{XX'} + \sqrt{1-q} \ket{11}_{XX'},
\end{equation}
where mode $X$ is kept at each party's site and mode $X'$ is sent to the central station. After the channel transmission, the quantum state comes to be as follows
\begin{equation}
    \ket{S}_{XE}\ket{1}_{X'} + \ket{V}_{XE} \ket{0}_{X'},
\end{equation}
where
\begin{equation}
    \begin{split}
        \ket{S} &= \sqrt{1-q} \sqrt{\eta} \ket{10}_{XE},\\
        \ket{V} &= \sqrt{q} \ket{00}_{XE} + \sqrt{1-q} \sqrt{1-\eta} \ket{11}_{XE}.
    \end{split}
\end{equation}
Here, $E$ denotes an environmental system corresponding to channel transmission, $\ket{S}$ corresponds to a state which inputs a single-photon to the interferometer $U$ and $\ket{V}$ corresponds to a state which inputs a vacuum state.

Next, we focus on the operation at the central station. For $N=4$, the operation of the interferometer can be expressed as
\begin{equation}\label{eq:trans4}
    \begin{split}
        U &= u^{\otimes 2} \\
        &= \frac{1}{2} \left( \begin{array}{cccc}
            1 & 1 & 1 & 1 \\
            -1 & 1 & -1 & 1 \\
            -1 & -1 & 1 & 1 \\
            1 & -1 & -1 & 1
        \end{array}  \right).
    \end{split}
\end{equation}
Let $D_1, D_2, D_3$ and $D_4$ denote the four single-photon detectors at the central station. We call events where the single-photon detector $D_1$ detects a single photon successful events. We can apply the same analysis for the situations where the other single-photon detectors detect a single photon. The measurement operation of the single-photon detector can be expressed as
\begin{equation}
    \{ (1-p_d) \ketbra{0}{0} , \ketbra{1}{1} + p_d \ketbra{0}{0}  \}.
\end{equation}
Here, we assume that the single-photon detectors can resolve the number of photons and discard events where two or more photons arrive at one detector. Then, the measurement operation where only the single-photon detector $D_1$ detects a single-photon is
\begin{equation}
    \begin{split}
        &(\ketbra{1}{1} + p_d \ketbra{0}{0})_{D_1} ((1-p_d) \ketbra{0}{0})_{D_2} \\
        &\otimes((1-p_d) \ketbra{0}{0})_{D_3} ((1-p_d) \ketbra{0}{0})_{D_4} \\
        = &(1-p_d)^3 \ketbra{1000}{1000} + p_d (1-p_d)^3 \ketbra{0000}{0000}.
    \end{split}
\end{equation}
We see that there are two measurement patterns, that is, the first term corresponds to events where dark counts do not occur and the second term corresponds to events where a dark count occurs at $D_1$. Let $\ket{\phi_1}$ and $\ket{\phi_2}$ denote distributed quantum states when the first pattern and the second pattern occur, respectively. From the transformation matrix of the interferometer (\ref{eq:trans4}), we can explictly express these two quantum states as follows
\begin{equation}
    \begin{split}
        \ket{\phi_1} &= \frac{1}{2} (\ket{SVVV} + \ket{VSVV} \\
        &\quad + \ket{VVSV} + \ket{VVVS})_{\boldsymbol{X}},
    \end{split}
\end{equation}
\begin{equation}
    \ket{\phi_2} = \ket{VVVV}_{\boldsymbol{X}}.
\end{equation}
Then, a joint quantum state among the parties $\rho$ is as follows
\begin{equation}\label{eq:distributedstate}
    \rho = \frac{1}{P} ((1-p_d)^3\ketbra{\phi_1}{\phi_1} + p_d (1-p_d)^3 \ketbra{\phi_2}{\phi_2}),
\end{equation}
where
\begin{align}
        P &= P_1+P_2\\
        P_1 &= \text{Tr} [(1-p_d)^3 \ketbra{\phi_1}{\phi_1}],\\
        P_2 &= \text{Tr} [p_d (1-p_d)^3 \ketbra{\phi_2}{\phi_2}].
\end{align}
Then, the success probability that successful events occur $P_\text{succ}$ is as follows
\begin{equation}
    P_\text{succ} = P.
\end{equation}

We can calculate a joint quantum state distributed over the legitimate parties for $N=3$ by considering that the input of the 4-th mode is vacuum. In this situation, we can similarly express $\ket{\phi_1}$ and $\ket{\phi_2}$ as
\begin{equation}
    \begin{split}
        \ket{\phi_1} &= \frac{1}{2} (\ket{SVV0} + \ket{VSV0}\\
        &\quad + \ket{VVS0} + \ket{VVV1})_{\boldsymbol{X}},
    \end{split}
\end{equation}
\begin{equation}
    \ket{\phi_2} = \ket{VVV0}_{\boldsymbol{X}}.
\end{equation}
Since the 4-th mode is vacuum, then,
\begin{align}
        \ket{\phi_1} &= \frac{1}{2} (\ket{SVV} + \ket{VSV} + \ket{VVS})_{\boldsymbol{X}},\\
        \ket{\phi_2} &= \ket{VVV}_{\boldsymbol{X}}.
\end{align}
By using these quantum states, we can express a quantum state distributed among the parties where $N=3$ as (\ref{eq:distributedstate}).

\section{Bell inequalities largely violated by \textit{W} states}
In this section, we describe Bell inequalities which we construct for the scenarios $(3, 3, 2)$, $(4, 2, 2)$ and $(5, 2, 2)$.

\subsection{(3,3,2) scenario}
For the $(3, 3, 2)$ scenario, we assume the parties perform the following measurements and solve the optimization problem.
\begin{alignat}{2}
    M_{0|0} &= 0.5 (I+\Pi(\pi/2)), & \, M_{1|0} &= I-M_{0|0},\\
    M_{0|1} &= 0.5 (I+\Pi(0)), & \, M_{1|1} &= I-M_{0|1},\\
    M_{0|2} &= 0.5 (I+\Pi(\pi/4)), & \, M_{1|2} &= I-M_{0|2},\\
    N_{0|0} &= 0.5 (I+\Pi(2\pi/3)), & \, N_{1|0} &= I-N_{0|0},\\
    N_{0|1} &= 0.5 (I+\Pi(\pi/3)), & \, N_{1|1} &= I-N_{0|1}\\
    N_{0|2} &= 0.5 (I+\Pi(0)), & \, N_{1|2} &= I-N_{0|2}.
\end{alignat}
We also assume that $\text{Bob}_1$ and $\text{Bob}_2$ perform the same measurements. Then, we get the following Bell expression.
\begin{widetext}
\begin{equation}
    \begin{alignedat}{13}
    [0&.247, &~  13&.080, &~  13&.080, &~ -72&.982, &~ -5&.801, &  -22&.874, &  -22&.874, &~  170&.097, &~ 76&.615, &  -43&.969,\\
     -47&.512, &~   6&.441,& -52&.987, &  14&.794, &   14&.381, &  36&.414, &~ -86&.001, &  214&.380, &   22&.513, &~  -93&.969,\\
       6&.937, &  -40&.572, &   -3&.913, &~   78&.416, & 76&.615, &~ -47&.512, & -43&.969,  &   6&.441,&-52&.987, &  14&.381,\\
         14&.794, &  36&.414,&~ 196&.613, & -97&.512,&  -97&.512, &  31&.031,&-100&.764, &  34&.582, &  34&.582, &  16&.530,\\
          -39&.867,  & 64&.520,  & -0&.277, & -35&.502,& 27&.964,&  -80&.522,  &-24&.228, & 204&.149,& -86&.001, &  22&.513,\\
           214&.380, & -93&.969, & 6&.937, &   -3&.913, & -40&.572, &  78&.416, & -39&.867, &  -0&.277,  & 64&.520, & -35&.502,\\
    27&.964, & -24&.228, & -80&.522,&  204&.149,& -30&.035,  & -7&.201, &  -7&.201, &   24&.816,& -55&.651, &  61&.994,\\
     61&.994, & -34&.047,& 92&.225, & -20&.682, & -20&.682, &  15&.662,& -104&.003, &  10&.995,  &  10&.995, &   6&.796,\\
      173&.331, &  -33&.981,&  -231&.123, & 186&.357,&-95&.361, &   7&.696, &  68&.780, & -36&.319, &-16&.427,&   88&.784,\\
       120&.053,& -142&.446, &-37&.728, & -17&.368, & -65&.610, & 116&.228,&  173&.331,& -231&.123,&  -33&.981, & 186&.357,\\
        -95&.361, &  68&.780, &   7&.696, &  -36&.319, & 24&.245, & -34&.600, & -34&.600, & 221&.081,&-9&.877, &   2&.065,\\
          2&.065, & -30&.735, &124&.730, &-168&.674, & -18&.698,&   98&.759,& -247&.755,&  130&.879, &   -4&.775, & -20&.071,\\
   -16&.427, & 120&.053, &  88&.784,& -142&.446,& -37&.728,&  -65&.610, & -17&.368, & 116&.228,&124&.730,&  -18&.698,\\
   -168&.674, &  98&.759,& -247&.755, &  -4&.775, & 130&.879, & -20&.071,& -29&.883, & 275&.227,&   275&.227,& -1154&.797, \\
   -54&.086,  &-40&.034,&  -40&.034, & 433&.671,& 47&.449, &   -17&.500, & -17&.500,&   -7&.461,& -55&.911,&    7&.990,\\
      7&.990, &  33&.021,& 99&.826,&  -42&.233, & -63&.291, &  24&.590,&-67&.393, &  13&.525,&   23&.663,  & 15&.686,\\
       -20&.685,  &  52&.625, &  27&.461,&  -69&.950,&  -32&.289,&   -6&.489,&   -7&.916, &   60&.800,&99&.826,&  -63&.291,\\
         -42&.233,&   24&.590,& -67&.393, &  23&.663, &  13&.525, &   15&.686, &106&.177, & -67&.836, & -67&.836,  & 43&.0166,\\
   -63&.472, &  22&.062, &  22&.062,  &  7&.759,& 20&.049,   &-8&.242, & -14&.055,  &  0&.605, &-30&.694,  &  1&.279,\\
     -9&.196, &   45&.204,&-20&.685, &  27&.461,&   52&.625,&  -69&.950,&  -32&.289,&   -7&.916, &  -6&.489, &  60&.800,\\
      20&.049, & -14&.055,&   -8&.242, &   0&.605,& -30&.694, &  -9&.196, &   1&.279, &  45&.204,&-43&.090,  & 53&.739,\\
         53&.739,&  -71&.156,&-39&.214, &  -2&.950,&   -2&.950,  & 54&.100 ]
    \end{alignedat}
\end{equation}
\end{widetext}

\subsection{(4,2,2) scenario}
For the $(4, 2, 2)$ scenario, we assume the parties perform the following measurements and solve the optimization problem.
\begin{alignat}{2}
    M_{0|0} &= 0.5 (I+\Pi(\pi/2)), & \, M_{1|0} &= I-M_{0|0},\\
    M_{0|1} &= 0.5 (I+\Pi(0)), & \, M_{1|1} &= I-M_{0|1},\\
    N_{0|0} &= 0.5 (I+\Pi(2\pi/3)), & \, N_{1|0} &= I-N_{0|0},\\
    N_{0|1} &= 0.5 (I+\Pi(\pi/3)), & \, N_{1|1} &= I-N_{0|1},
\end{alignat}
where we assume that $\text{Bob}_1$, $\text{Bob}_2$ and $\text{Bob}_3$ perform the same measurements on the \textit{W} state $\ket{W_4}$. Then, we get the following Bell expression.
\begin{widetext}
\begin{equation}
    \begin{alignedat}{13}
        [-175&.500, & 157&.364, & 157&.364, & -61&.255, &157&.364,  &-61&.255, & -61&.255, & -81&.627,  &31&.906, & -45&.094, \\
        -45&.094, & -21&.089, & -45&.094, & -21&.089,&  -21&.089,&  395&.608, & -20&.805,  &  13&.390,&  182&.787, & -74&.300,\\
        182&.787, & -74&.300,& -440&.494, &  55&.609, & 84&.976,& -102&.482,& -167&.082,&   62&.336, &-167&.082, &  62&.336, \\
          92&.223, &  84&.100, &-20&.805,&  182&.787,&   13&.390,  & -74&.300, &182&.787,& -440&.494, & -74&.300, &  55&.609,\\
          84&.976,& -167&.082,& -102&.482,&   62&.336, &-167&.082, &  92&.223, &  62&.336, &  84&.100, &76&.176, &  42&.004,\\
           42&.004,& -113&.982,  &84&.863,& -186&.739, & -186&.739,&  150&.889,  &70&.580, &-143&.281, &-143&.281, &  93&.027,\\
  -419&.395, & 211&.812, & 211&.812,  &  -5&.681, & -20&.805,  & 182&.787, & 182&.787,& -440&.494,  &13&.390, & -74&.300,\\
  -74&.300,&   55&.609,&   84&.976,& -167&.082,& -167&.082, &  92&.223,&-102&.482,&    62&.336, &  62&.336,  & 84&.100, \\
  76&.176,  &  42&.004,  & 84&.863,& -186&.739,& 42&.004, & -113&.982,& -186&.739,&  150&.889,   &70&.580,& -143&.281, \\
   -419&.396, & 211&.812, & -143&.281, &   93&.027, &  211&.812,  & -5&.681,& 76&.176, &  84&.863, &  42&.004,& -186&.739,\\
   42&.004,& -186&.739, &-113&.982,&  150&.889, &70&.580, &-419&.395,& -143&.281, & 211&.812, & -143&.281, & 211&.812,\\
    93&.027,  &  -5&.681,  &514&.847, &  -45&.192,&  -45&.192, &  -48&.576, &-45&.192,&  -48&.576, & -48&.576, &  72&.284,\\
   -36&.318, & -91&.061,&  -91&.061, & 136&.543, &-91&.061,&  136&.543,&   136&.543,&  -32&.732, & -415&.077,&  152&.143,\\
   152&.143, &-222&.524,& 152&.143, &-222&.524,& -222&.524, &  94&.050, &97&.836,&  -42&.045,&  -42&.045,&   69&.817, \\
   -42&.045, &  69&.817,&   69&.817, &  22&.189, & 9&.454, & -80&.168,&   26&.822,&  -56&.515,  &26&.822,&  -56&.515, \\
   -735&.028, & 226&.965,  &48&.963, &  -2&.637, &   -20&.330,  & 45&.615,&-20&.330, &  45&.615,  &   125&.958, & -16&.209,\\
    9&.454, &  26&.822, &  -80&.168, & -56&.515, &26&.822, &-735&.028, & -56&.515, & 226&.965,& 48&.963, & -20&.330,\\
     -2&.637,&   45&.615, &-20&.330,&  125&.958, &    45&.615,&  -16&.209, &350&.080, &  -196&.220,& -196&.220, &    8&.327,\\
 -1257&.239, &  43&.335, &  43&.335,  & 51&.046, &-37&.404,&   68&.115, &  68&.115,&  -22&.020,  &151&.249,&  -33&.866,\\
  -33&.866,  & 65&.404,&9&.454, &  26&.822, &  26&.822,& -735&.028,  &-80&.168, & -56&.515, & -56&.515,&  226&.965,\\
  48&.963, & -20&.330, & -20&.330, & 125&.958,&    -2&.637, &  45&.615,  &  45&.615, & -16&.209, &350&.080,& -196&.220, \\
  -1257&.239, &  43&.335,&-196&.220,&    8&.327,&   43&.335, &  51&.046,&  -37&.404, &  68&.115, & 151&.249,  &   -33&.866,\\
  68&.115,&  -22&.020,&  -33&.866, &  65&.404, & 350&.080,& -1257&.239,& -196&.220, &  43&.335, &-196&.220, &  43&.335,\\
   8&.327, &  51&.046, &-37&.404,&  151&.249, &  68&.115, & -33&.866, & 68&.115, & -33&.866, & -22&.020, &  65&.404,\\
   307&.995,& -915&.232, &-915&.232, & 160&.203, &-915&.232,&  160&.203, & 160&.203, & -67&.435,  &-8&.768, & 102&.174, \\
    102&.174, & -61&.625,&102&.174,&  -61&.625, & -61&.625, &  116&.453]
    \end{alignedat}
\end{equation}
\end{widetext}

\subsection{(5,2,2) scenario}
For the $(5, 2, 2)$ scenario, we assume the parties perform the following measurements and solve the optimization problem.
\begin{alignat}{2}
    M_{0|0} &= 0.5 (I+\Pi(\pi/2)), & \, M_{1|0} &= I-M_{0|0},\\
    M_{0|1} &= 0.5 (I+\Pi(0)), & \, M_{1|1} &= I-M_{0|1},\\
    N_{0|0} &= 0.5 (I+\Pi(2\pi/3)), & \, N_{1|0} &= I-N_{0|0},\\
    N_{0|1} &= 0.5 (I+\Pi(\pi/3)), & \, N_{1|1} &= I-N_{0|1}.
\end{alignat}
Same as the other scenarios, we assume that Bobs perform the same measurements on the \textit{W} state $\ket{W_5}$.

\section{State calculation with Gaussian formalism}\label{appendix:Gaussian}
We calculate joint probability distribution of legitimate parties. Here, we give an argument for $N=3, 4$.

First, denote $N$ pairs of anihilation and creation operators $\{\hat{a}_i, \hat{a}^\dagger_i  \}_{i=1}^N$ which satisfy the following commutation relations
\begin{equation}
    [\hat{a}_i, \hat{a}^\dagger_j] = \delta_{ij}.
\end{equation}
By using these operators, we define the quadrature field operators $\{\hat{q}_i, \hat{p}_i  \}_{i=1}^N$ as follows
\begin{align}
    \hat{q}_i &= \hat{a}_i + \hat{a}^\dagger_i,\\
    \hat{p}_i &= i(\hat{a}^\dagger_i -\hat{a}_i ).
\end{align}
In the $N$-mode bosonic system, a quantum state $\rho$ is characterized by its characteristic function
\begin{equation}
    \chi(x) = \text{Tr} [ \rho \mathcal{W} (x)],
\end{equation}
where $\mathcal{W}$ is a Weyl operator defined as
\begin{equation}
    \mathcal{W} (x) = \exp [-i x^T r],
\end{equation}
$x \in \mathbb{R}^{2N}$ is a real vector, and $r = [\hat{q}_1, \hat{p}_1, \ldots, \hat{q}_N, \hat{p}_N]^T$ is a vector that consists of the qudrature operators.

The covariance matrix of a two-mode squeezed state with mean photon number $\bar{n}$ is as follows
\begin{equation}
    \Gamma^{\text{TMSV}} (\bar{n}) = \left[ \begin{array}{cc}
    v I_2 & \sqrt{v^2-1} Z\\
    \sqrt{v^2-1} Z & v I_2
    \end{array}
    \right],
\end{equation}
where $I_n$ is the $n \times n$ identity matrix and
\begin{align}
    Z &= \left[  
    \begin{array}{cc}
        1 & 0 \\
        0 & -1
    \end{array}
    \right].
\end{align}
The parameter $v$ corresponds to squeezing level of the two-mode squeezed state, and it relates to the mean photon number of the state $\bar{n}$ as $v = 1 + 2\bar{n}$. Here, we omit an argument for a displacement vector since it is always a zero vector. Then, a coveriance matrix of an initial state is as follows
\begin{equation}
    \Gamma^{\text{initial}} = \Gamma_{\boldsymbol{X} \boldsymbol{X'}}^{\text{TMSV}} (\bar{n}) \oplus I_{\boldsymbol{E} \boldsymbol{F}},
\end{equation}
where
\begin{equation}
\begin{split}
    \Gamma_{\boldsymbol{X} \boldsymbol{X'}}^{\text{TMSV}} (\bar{n}) &= \Gamma_{AA'}^{\text{TMSV}} (\bar{n}) \oplus \Gamma_{B_1B_1'}^{\text{TMSV}} (\bar{n})\\
    &\quad \oplus \Gamma^{\text{TMSV}}_{B_2B_2'} (\bar{n}) \oplus \Gamma^\text{TMSV}_{B_3B_3'} (\bar{n}),
\end{split}
\end{equation}
modes $A$ and $B_i$ are detected by Alice and $\text{Bob}_i$ for $i \in \{ 1, \ldots, N-1 \}$, modes $A'$ and $B_i'$ are sent to the central station, $\boldsymbol{E}$ denotes environmental modes for the pure-loss channels, and $\boldsymbol{F}$ denotes environmental modes for the on-off detectors of the parties, respectively. We can model a pure-loss channel with transmissivity $\eta$ as a beamsplitter with vacuum in one of the input modes. Also, we can effectively shift the detection efficiencies of the on-off detectors at each party's sites prior to their displacement operations since any loss at the detectors can be compensated by changing the amplitude. The operation of the beamsplitter with transmissivity $\eta$ can be expressed as follows
\begin{equation}
    S^{\text{BS}} (\eta) = \left[
    \begin{array}{cc}
        \sqrt{\eta}\, I_2 & \sqrt{1-\eta}\, I_2 \\
        -\sqrt{1-\eta}\, I_2 & \sqrt{\eta}\, I_2
    \end{array}
    \right].
\end{equation}
We can obtain a state after the channel transmission by applying this operation to the modes denoted by $\boldsymbol{X'}$ and $\boldsymbol{E}$. 
Next, we consider the operation of the central node. The operation of a 50:50 beamsplitter is
\begin{equation}
    S^{\text{50:50BS}} = S^{\text{BS}} \left( \frac{1}{2} \right) = \frac{1}{\sqrt{2}} \left[ 
    \begin{array}{cc}
        I_2 & I_2 \\
        -I_2 & I_2
    \end{array}
    \right].
\end{equation}
Then, we can calculate a quantum state after the interferometer $U$ by appropriately applying this operation to the quantum state. We consider detection efficiencies of the on-off detectors at each party's sites. These operations are expressed by beamsplitters with transmissivity $\eta_e$. Therefore, by applying these operations, we can get a quantum state before the single-photon interference, and let $\sigma$ and $\Gamma^\sigma$ denote this state and its covariance matrix, respectively.

Next, we consider the detection at the central node. We assume that events where the detector of the mode $A'$ clicks and those of the other modes do not click. The POVM corresponding to the on-off detector is 
\begin{equation}
    \{ (1-p_{d}) \ketbra{0}{0}, I-(1-p_{d}) \ketbra{0}{0}\},
\end{equation}
where $I$ is the identity operator. Then, a joint quantum state between the parties conditioned on successful heralding is
\begin{equation}
\begin{split}
    \rho_{\boldsymbol{X}} 
    &=q_1 \rho_{\boldsymbol{X}}^{(1)} - q_2 \rho_{\boldsymbol{X}}^{(2)},
\end{split}
\end{equation}
where
\begin{align}
    \rho_{\boldsymbol{X}}^{(1)} &= \tilde{\rho}_{\boldsymbol{X}}^{(1)}/\text{Tr} [\tilde{\rho}_{\boldsymbol{X}}^{(1)}],\\
    \rho_{\boldsymbol{X}}^{(2)} &= \tilde{\rho}_{\boldsymbol{X}}^{(2)}/\text{Tr} [\tilde{\rho}_{\boldsymbol{X}}^{(2)}],\\
    \tilde{\rho}_{\boldsymbol{X}}^{(1)} &= \text{Tr}_{\boldsymbol{X'} \boldsymbol{E} \boldsymbol{F}} [ \sigma \ketbra{000}{000}_{\boldsymbol{X'}/A'}   ],\\
    \tilde{\rho}_{\boldsymbol{X}}^{(2)} &= \text{Tr}_{\boldsymbol{X'} \boldsymbol{E} \boldsymbol{F}} [ \sigma \ketbra{0000}{0000}_{\boldsymbol{X'}}    ],\\
    q_1 &= \frac{(1-p_{d})^3\text{Tr}[\sigma \ketbra{000}{000}_{\boldsymbol{X'}/A'}]}{P},\\
    q_2 &= \frac{(1-p_{d})^4 \text{Tr}[\sigma \ketbra{0000}{0000}_{\boldsymbol{X'}} ]}{P},
\end{align}
and $P$ is a probability that successful events occur;
\begin{equation}
\begin{split}
    P &= (1-p_{d})^3 \text{Tr} [\sigma \ketbra{000}{000}_{\boldsymbol{X'}/A'}]\\
    &\quad - (1-p_d)^4 \text{Tr} [\sigma \ketbra{0000}{0000}_{\boldsymbol{X'}}].
\end{split}
\end{equation}
Here, $\boldsymbol{X'}/A'$ means the modes $B_1' B_2' B_3'$.

By tracing out the modes except for $\boldsymbol{X}'/A'$, and $\boldsymbol{X'}$ from $\Gamma^{\sigma}$, we obtain covariance matrices $\Gamma_{\boldsymbol{X}'/A'}$ and $\Gamma_{\boldsymbol{X}'}$, respectively. By using these covariance matrices, we calculate the following probabilities
\begin{equation}
    \begin{split}
        \text{Tr} (\sigma \ketbra{000}{000}_{\boldsymbol{X}'/A'}) &= \frac{8}{\sqrt{\det\{ \Omega_6 (\Gamma_{\boldsymbol{X}'/A'} + I_6) \Omega_6^T \}}},\\
        \text{Tr} (\sigma \ketbra{0000}{0000}_{\boldsymbol{X}'}) &= \frac{16}{\sqrt{\det\{ \Omega_8 (\Gamma_{\boldsymbol{X}'} + I_8) \Omega_8^T \}}},\\        
    \end{split}
\end{equation}
where $\Omega_n$ is defined as follows
\begin{equation}
    \Omega_n = \bigoplus_{k=1}^{n/2} \omega_k = \left( 
    \begin{array}{ccc}
        \omega &  &  \\
         & \ddots & \\
         &  & \omega
    \end{array}
    \right), \omega = \left[\begin{array}{cc}
        0 & 1 \\
       -1  & 0
    \end{array} \right]
\end{equation}

Then, we calculate the probability distribution $p(a,\boldsymbol{b}|x,\boldsymbol{y})$. To do this, we derive covariance matrices $\Gamma^{(1)}$ and $\Gamma^{(2)}$ which corresponding to $\rho_{\boldsymbol{X}}^{(1)}$ and $\rho_{\boldsymbol{X}}^{(2)}$, respectively. Let $\Gamma_{AA'B_1B_1'B_2B_2'B_3B_3'}$ denote the covariance matrix obtained by tracing out the environmental modes from $\Gamma^{\sigma}$. By changing the ordering of the modes, we obtain $\Gamma_{AB_1B_2B_3A'B_1'B_2'B_3'}$, and this covariance matrix has a following form
\begin{equation}
    \Gamma_{AB_1B_2B_3A'B_1'B_2'B_3'} = \left[ 
    \begin{array}{cc}
        \Gamma_{AB_1B_2B_3A'B_1'B_2'} & C_{B_3'} \\
        C_{B_3'}^T & \Gamma_{B_3'}
    \end{array}
    \right].
\end{equation}
Since the projection on vacuum can be considered as heterodyne measurement, a covariance matrix after the projection on vacuum on the mode $B_3'$ is
\begin{equation}
\begin{split}
        \Gamma'_{AB_1B_2B_3A'B_1'B_2'} &= \Gamma_{AB_1B_2B_3A'B_1'B_2'}\\
        &\quad- C_{B_3'} (\Gamma_{B_3'} + I_2)^{-1} C_{B_3'}^T\\
        &= \left[ 
        \begin{array}{cc}
            \Gamma_{AB_1B_2B_3A'B_1'} & C_{B_2'} \\
            C_{B_2'}^T & \Gamma_{B_2'}
        \end{array}
        \right].
\end{split}
\end{equation}
By repeating this procedure, we get $\Gamma^{(1)}$ and $\Gamma^{(2)}$.
% \begin{equation}
% \begin{split}
%         \Gamma'_{ABCDA'B'} &= \Gamma_{ABCDA'B'}- C_{C'} (\Gamma_{C'} + I_2)^{-1} C_{C'}^T\\
%         &= \left[ 
%         \begin{array}{cc}
%             \Gamma_{ABCDA'} & C_{B'} \\
%             C_{B'}^T & \Gamma_{B'}
%         \end{array}
%         \right].
% \end{split}
% \end{equation}
% \begin{equation}
% \begin{split}
%         \Gamma'_{ABCDA'} &= \Gamma_{ABCDA'}- C_{B'} (\Gamma_{B'} + I_2)^{-1} C_{B'}^T\\
%         &= \left[ 
%         \begin{array}{cc}
%             \Gamma_{ABCD} & C_{A'} \\
%             C_{A'}^T & \Gamma_{A'}
%         \end{array}
%         \right].
% \end{split}
% \end{equation}
% Therefore, $\Gamma^{(1)}$ and $\Gamma^{(2)}$ can be derived as follows
% \begin{align}
%         \Gamma^{(1)} &= \text{Tr}_{A'} \left[ \Gamma_{ABCDA'} \right] = \Gamma_{ABCD},\\
%         \Gamma^{(2)} &= \Gamma_{ABCD} - C_{A'} (\Gamma_{A'}+I_2)^{-1} C_{A'}^T.
% \end{align}
Let $\alpha$, $\beta$, $\gamma$ and $\delta$ be displacements of the parties, respectively. Then, after performing these displacement operations, a displacement vector of the joint quantum state among the parties comes to be
% \begin{equation}
%     d_{\boldsymbol{X}} = 2\left[   \begin{array}{c}
%         \text{Re} (\alpha)  \\
%          \text{Im} (\alpha) \\
%          \text{Re} (\beta) \\
%          \text{Im} (\beta) \\
%          \text{Re} (\gamma) \\
%          \text{Im} (\gamma) \\
%          \text{Re} (\delta) \\
%          \text{Im} (\delta)
%     \end{array} \right].
% \end{equation}
\begin{equation}
\begin{split}
    d_{\boldsymbol{X}} = 2[&\text{Re} (\alpha), \text{Im} (\alpha), \text{Re} (\beta), \text{Im} (\beta), \\
         &\text{Re} (\gamma), \text{Im} (\gamma), \text{Re} (\delta), \text{Im} (\delta) ]^T.
\end{split}
\end{equation}
We calculate a probability $p(0,\ldots,0|x,\boldsymbol{y})$ where all of the parties on-off detectors do not click. POVM of these detectors can be expressed as
\begin{equation}
    \{ \ketbra{0}{0} , I - \ketbra{0}{0}  \}.
\end{equation}
Then, we calculate this probability in the following way
\begin{equation}
\begin{split}
    &p(0,\ldots,0|x,\boldsymbol{y})\\
    &\quad = \text{Tr} \left[ \rho_{\boldsymbol{X}} \ketbra{0000}{0000}_{\boldsymbol{X}} \right] \\
    &\quad=\text{Tr} \left[q_1 \rho_{\boldsymbol{X}}^{(1)} \ketbra{0000}{0000}_{\boldsymbol{X}}  \right]\\
    &\quad \quad - \text{Tr} \left[q_2 \rho_{\boldsymbol{X}}^{(2)} \ketbra{0000}{0000}_{\boldsymbol{X}}   \right].
\end{split}
\end{equation}
Here, let $\chi^{(1)}(x)$ be a characteristic function of $\rho^{(1)}_{\boldsymbol{X}}$ and $\chi_0(x)$ be a characteristic function of projection onto vacuum defined as follows
\begin{align}
        \chi^{(1)} (x) &= \exp (\frac{1}{2} x^T (\Omega_8 \Gamma^{(1)} \Omega_8^T) x -i (\Omega_8 d_{\boldsymbol{X}})^T x), \\
        \chi_0 (x) &= \exp(-\frac{1}{2} x^T (\Omega_8 I_8 \Omega_8^T) x),
\end{align}
where $x \in \mathbb{R}^8$ is a real vector. Then, the following holds
\begin{widetext}
\begin{equation}
    \begin{split}
        \text{Tr} \left[q_1 \rho_{\boldsymbol{X}}^{(1)} \ketbra{0000}{0000}_{\boldsymbol{X}} \right] &= \frac{q_1}{\pi^2} \int \chi^{(1)} (x) \chi_0 (x) \, dx \\
        &= \frac{16 q_1}{\sqrt{\det (\Omega_8 (\Gamma^{(1)} + I_8) \Omega_8^T )} } \exp (-\frac{1}{2} (\Omega_8 d_{\boldsymbol{X}})^T (\Omega_8 (\Gamma^{(1)} + I_8) \Omega_8^T)^{-1} (\Omega_8 d_{\boldsymbol{X}})) \\
    \end{split}
\end{equation}
\end{widetext}
We can analogously calculate the term $\text{Tr} \left[q_2 \rho_{\boldsymbol{X}}^{(2)} \ketbra{0000}{0000}_{\boldsymbol{X}}   \right]$, and we can derive the probability $p(0,0,0,0|x,\boldsymbol{y})$. By applying the same procedure, we can calculate the other probabilities.

\bibliography{WDIQKD}% Produces the bibliography via BibTeX.

\end{document}